\documentclass[12pt]{article}
\usepackage{amssymb,amsmath,slashed}
\usepackage[pdftex]{graphicx}

\makeatletter

\@addtoreset{equation}{section}
\def\section{\@startsection {section}{1}{\z@}{-2.5ex plus -1ex minus
 -.2ex}{1.3ex plus .2ex}{\large\bf}}
\def\subsection{\@startsection{subsection}{2}{\z@}{-2.25ex plus%
 -1ex minus -.2ex}{0.5ex plus .2ex}{\bf}}

\advance \voffset by -1.0in
\advance \hoffset by -0.7in
\def\sec{f}
\def\Sec{F}
\def\gf{\mathcal F}
\def\gp{\mathcal A}
\def\lf{\sigma}
\def\lv{X}
\def\rv{Z}
\def\Dslash{\slashed{D}}
\def\bee{\begin{equation}}
\def\eee{\end{equation}}
\def\bea{\begin{align}}
\def\eea{\end{align}}

\newcommand{\CP}{\mathbb{CP}}

\def\bone{{\mbox{\bf 1}}_2}

\def\bpm{\begin{pmatrix}}
\def\epm{\end{pmatrix}}

\newcommand{\NN}{\mathbb{N}}
\newcommand{\ZZ}{\mathbb{Z}}

\newcommand{\RR}{\mathbb{R}}
\newcommand{\CC}{\mathbb{C}}

\def\bee{\begin{equation}}
\def\eee{\end{equation}}

\def\cA{\mathcal A}
\def\bea{\begin{align}}
\def\eea{\end{align}}

\begin{document}
\begin{flushright}
EMPG-13-23
\end{flushright}
\vskip 10pt
\baselineskip 28pt

\begin{center}
{\Large \bf  Dirac operators on the Taub-NUT space, monopoles  \\ and $SU(2)$ representations}

\baselineskip 18pt

\vspace{1 cm}

{\bf Rogelio Jante and Bernd J.~Schroers},\\
\vspace{0.2 cm}
Maxwell Institute for Mathematical Sciences and
Department of Mathematics,
\\Heriot-Watt University,
Edinburgh EH14 4AS, UK. \\
{\tt rj89@hw.ac.uk} and {\tt b.j.schroers@hw.ac.uk} \\
\vspace{0.4cm}

{ December 2013} 
\end{center}

\begin{abstract}
\noindent
We  analyse the normalisable zero-modes of the Dirac operator on the Taub-NUT manifold coupled to an abelian gauge field with self-dual curvature,  and interpret them in terms of the zero modes of the Dirac operator on the 2-sphere coupled to a Dirac monopole. We show that the space of zero modes decomposes into a direct sum of irreducible $SU(2)$ representations of all dimensions up to a bound determined by the  spinor charge  with respect to the abelian gauge group.  Our decomposition provides  an interpretation of an index formula  due to Pope and provides a possible model for spin in recently proposed geometric models of matter.

\end{abstract}

\baselineskip 16pt
\parskip 5 pt
\parindent 0pt

\section{Introduction}
\subsection{Motivation and overview of main results}

The Dirac equation on the 2-sphere and coupled to a Dirac monopole provides one of the simplest illustrations of an index theorem \cite{AS}. For a monopole of magnetic charge $g$ and a spinor of electric charge $e$, the product of electric and magnetic charge is an integer multiple of Planck's constant by Dirac's quantisation condition, i.e., 
\bee
\label{Dirac}
\frac{eg}{2\pi \hbar}=n\in \ZZ.
\eee
 In mathematical terms,  coupling to a Dirac monopole amounts to twisting the  Dirac operator on the 2-sphere by a complex line bundle with connection. The integer  $n$ is the Chern number of that line bundle and the index of the twisted Dirac operator turns out to be  $n$, too. Together with a vanishing theorem, this gives the dimension of the space of zero modes as $|n|$, see e.g. \cite{Varilly} and \cite{ES} for recent treatments and reviews. In physical terms, there is therefore one state per cell of volume $2\pi \hbar$ in the electric-magnetic charge plane. 

The index is independent of the detailed form of the magnetic field and the metric on the 2-sphere. However, by specialising to the round metric on the 2-sphere and the  rotationally invariant magnetic monopole  field,  we can bring the double cover  $SU(2)$ of the isometry group into the picture. The twisted Dirac operator and its kernel are now naturally acted on by  $SU(2)$ and the kernel is, in fact,  the irreducible $SU(2)$ representation of dimension $|n|$. Parametrising the 2-sphere in terms of a complex coordinate via stereographic projection, one can realise the zero modes in terms of holomorphic (for $n>0$) or antiholomorphic (for $n<0$) polynomials of degree $|n|-1$.

In this paper we will review these results and use them to gain a better understanding of an index formula due to Pope for the Dirac operator   on the Taub-NUT manifold, coupled to an abelian connection.  The Taub-NUT manifold is the static part of the Kaluza-Klein description of a magnetic monopole \cite{Sorkin,GP}.  It is a  Riemannian 4-manifold with a self-dual Riemann curvature and
has the structure of a circle bundle over $\RR^3\setminus \{0\}$, with the fibre collapsing at the origin. The geometry  encodes the Dirac monopole connection on this bundle away from the origin but is smooth even when the fibre shrinks to a point. In that sense, the situation we consider  may be thought of as a geometric and non-singular version of the Dirac operator coupled to a Dirac monopole on $\RR^3$.

 Topologically, the Taub-NUT manifold  is  $\CC^2$, and index theorems are generally more difficult on non-compact spaces. However, exploiting the explicit form and $U(2)$ symmetry of the Taub-NUT metric, Pope found  that,  after coupling to an abelian gauge field with a suitably defined flux $p$, the dimension of the kernel of the twisted Dirac operator $\Dslash_p$ on  Taub-NUT is
\bee
\label{popedim}
\text{dim ker}\, \Dslash_p = \frac 1 2 [|p|] ([|p|]+1),
\eee  
where, for a positive real number $x$,  we define $[x]$ as the largest integer {\em strictly} smaller than $x$ \cite{Pope1, Pope2}.  Here, we would like to understand the  $SU(2)$ transformation properties of these zero-modes, and we would like to gain a qualitative understanding why  the Dirac operator on Taub-NUT  only has  zero-modes if one twists it  by a further abelian gauge field - even though the  Taub-NUT geometry already encodes a Dirac monopole.

The curvature of the  gauge field considered by Pope is the, up to scale, unique rotationally symmetric, closed and self-dual 2-form on the Taub-NUT manifold with a finite $L^2$-norm. Since the Taub-NUT manifold is topologically trivial there is no natural normalisation of this form, but in our discussion we will fix the scale by normalising the integral over the `2-sphere at spatial infinity'.  In terms of the detailed discussion of the  Taub-NUT space  in \cite{AMS}, we normalise  the 2-form to be the Poincar\'e dual of the $\CP^1$ which compactifies  the Taub-NUT manifold  to $\CP^2$. 

With our  normalisation, we treat the 2-form as the curvature of a  (topologically  trivial) bundle over Taub-NUT. However, we allow   the structure group of the bundle   to be $(\RR,+)$ rather than $U(1)$ so that unitary representations  of  an element $u\in \RR$ are by a phase $e^{ipu}$ with $p\in \RR$. When we  twist the Dirac operator with this bundle, spinors may therefore have any {\em real} charge $p$.  On the topologically trivial Taub-NUT manifold,  there is no Dirac condition like \eqref{Dirac} to force the product of the `magnetic'  and `electric' charge to be an integer or, equivalently, the gauge group to be $U(1)$.  

 Here and in the rest of the paper we reserve electric-magnetic terminology for the $U(1)$-gauge field encoded in the geometry of Taub-NUT  and put  it in inverted commas  for the  auxiliary $\RR$-gauge field, as above. While the `electric' charge of spinors is  the external parameter $p$,   the electric charge of spinors is determined by the eigenvalue of the central $U(1)$ in the $U(2)$ isometry group. 
 We  find that the interplay between the two charges   determines the number of  normalisable Dirac zero-modes. 
 Assuming for simplicity $p>0$, we find that zero-modes are normalisable only if their electric charge satisfies   \eqref{Dirac} with    $n\leq [p]$.  Moreover, we learn that,  for each allowed value of $n$, there is an $n$-dimensional space of zero-modes, forming   an irreducible $SU(2)$ representations as for the Dirac monopole.  The space of zero-modes is the direct sum of these irreducible representations, reproducing and interpreting  Pope's dimension formula as the sum $1+2+\ldots +([p]-1)+[p]$.

Our interest in the zero-modes of the Dirac operator on the Taub-NUT manifold was triggered by  geometric models of elementary particles recently proposed in  \cite{AMS}.   In this framework,  the Taub-NUT manifold is a  model for the electron, and the zero-modes discussed in this paper are candidates for describing the spin degrees of the freedom of the electron. Our discussion shows that it is indeed possible to obtain a  spin 1/2 doublet of states from the  normalisable zero modes by picking $ 2 <p \leq 3$. However, with this choice one inevitably also obtains a spin 0 singlet, as $[p]$ only sets an upper limit on the dimensions of irreducible $SU(2)$ representations. We discuss possible interpretations of the doublet and the singlet at the end of our paper.

In view of the obvious generalisations of the Dirac operator studied here - for example to the 4-geometries with line bundles proposed as  geometric models for the proton and the neutron in \cite{AMS} - we have used this paper to prepare the ground for  studies along these lines. We  have taken care to  set up  consistent conventions  regarding the various line bundles, connections and $SU(2)$ actions which we use. In particular, we have found complex coordinates more convenient than the more widely used polar coordinates and Euler angles since the zero-modes can then be given in terms of  holomorphic sections of the relevant line bundles.

The paper is organised as follows. A brief summary of important background and conventions is given in the second half of this introduction, with much more detail provided in the Appendix. In Sect.~2 we review the zero-modes of the  Dirac operator coupled to the Dirac monopole, first on the 2-sphere and then on $\RR^3$ with a suitable mass term,  induced by dimensional reduction.  Sect.~3 treats the twisted Dirac operator on Taub-NUT, using the insights and terminology of Sect.~2. In view of possible extensions of our results we begin in a more general setting of self-dual and rotationally symmetric 4-manifolds but then specialise to the Taub-NUT manifold and the $\RR$-connection with a self-dual and normalisable curvature.  Sect.~4 contains our discussion and conclusions. 

\subsection{Conventions}

The Hopf fibration of the 3-sphere,  associated line bundles over the 2-sphere and various differential operators acting on their sections all  play important  roles in this paper. These are mostly standard topics but since we draw on a broad range of them - from harmonic analysis on $S^3$ to holomorphic sections of powers of the hyperplane bundle $H$ - we require a set of consistent conventions for the calculations in this paper. We have collected basic definitions and our conventions in  the  extended Appendix. It is explained there that  $H^n$ is the line bundle associated to the Lens space $L(1,n)$ and that the Dirac monopole of charge $n$ is an $SU(2)$-invariant $U(1)$ connection    on this bundle, with $n$ being both the monopole charge and the Chern number. Useful references for this material and its relation to Dirac operators are the papers \cite{Varilly, Dray1, Dray2} as well as, at a more introductory level, the textbooks \cite{GS,Sternberg}.

In the following discussions, we use both Euler angles  $(\alpha,\beta,\gamma)$ and complex coordinates $(z_1,z_2)$ with $|z_1|^2+|z_2|^2=1$ to parametrise $S^3\cong SU(2)$. Both are defined in  Appendix~\ref{parametrisations} and related via
\bee
\label{zdef}
z_1= e^{-\frac i 2 (\alpha+ \gamma)} 
\cos \frac{\beta}{2} , \quad 
z_2=
e^{\frac i 2 (\alpha - \gamma)} \sin \frac{\beta }{2}.
\eee
In angular coordinates, the Hopf map $S^3\rightarrow S^2$ maps $(\alpha,\beta,\gamma)$ to standard spherical polar coordinates $(\beta, \alpha)\in [0,\pi]\times [0,2\pi)$ on the 2-sphere. In this paper we mostly work with complex coordinates for the 2-sphere, with $z\in \CC$ parametrising  a northern patch $U_N$ (covering all but the South Pole) via stereographic projection from the South Pole, and $\zeta\in\CC$  parametrising  a southern patch   $U_S$ (covering all but the North Pole) via stereographic projection  from the North Pole and complex conjugation.  The details are in 
Appendix~\ref{Hopfconventions}, which also  includes definitions of local sections $s_N:U_N\rightarrow S^3$ and 
$s_S:U_S\rightarrow S^3$. The resulting relation between complex and angular coordinates is
\bee 
\label{zdefined}
z=\frac{z_2}{z_1} = \tan\frac{\beta}{2} e^{i\alpha} , \qquad 
\zeta=\frac{z_1}{ z_2} =\cot\frac{\beta}{2} e^{-i\alpha}.
\eee

The left-invariant 1-forms $\sigma_1,\sigma_2$ and $\sigma_3$ on $SU(2)$ are important in this paper and are defined and expressed in terms of the Euler angles and complex coordinates in  Appendix~\ref{formsfields}.  The dual  left-invariant (and right-generated) vector fields $\lv_1,\lv_2$ and $\lv_3$ are also defined and evaluated there.  For our discussion of the monopoles we need in particular the expression for the 1-form
\bee
\lf_3= d\gamma + \cos\beta d\alpha = 2i(\bar{z}_1 dz_1 + \bar{z}_2dz_2)
\eee
and the dual vector field
\bee
\label{x3defined}
\lv_3=\partial_\gamma =\frac  i  2 ( \bar{z}_1\bar{\partial}_1 +\bar{z}_2 \bar{\partial}_2 - z_1 \partial_1 -z_2 \partial_2). 
\eee

Finally, our conventions regarding the Dirac operator  on Riemannian manifold  are collected in Appendix~\ref{Diracconventions}.  Generally, when working with  numbered local coordinates $x_1,\ldots,x_n$ we write $\partial_1,\ldots,\partial_n$ for the associated partial derivatives. When working with  alphabetically named coordinates $\alpha, \beta,\gamma \ldots$ we write $\partial_\alpha,
\partial_\beta,\partial_\gamma \ldots$ for the associated partial derivatives. We use the Einstein summation convention throughout.

\section{The Dirac operator coupled to the Dirac monopole}

\subsection{Twisted Dirac operators on the  2-sphere}
We review the  the Dirac operator on the unit 2-sphere, with its round metric.  In terms of spherical coordinates $(\beta,\alpha)\in [0,\pi]\times [0,2\pi)$ the line element is
\bee
ds^2 = d\beta^2 + \sin^2\beta d\alpha^2,
\eee
so that we could work with 2-bein $
{\tilde e}_1 = d\beta, \, {\tilde e}_2= \sin\beta d\alpha $,
and the associated frame
\bee
\label{angframe}
{\tilde E}_1 =\partial_\beta, \qquad {\tilde E}_2=\frac{1}{\sin\beta} \partial_\alpha.
\eee
This frame has the disadvantage of being ill-defined on both the North and the South Pole. 
In terms of the complex coordinate  $z$ \eqref{zdefined}, which is defined everywhere but at the South Pole of  $S^2$, 
 the metric reads
\bee
ds^2 = \frac  4  {q^2} dzd\bar{z},
\eee
where 
\bee
\label{qdef}
q = 1 + z\bar{z}.
\eee
Writing $z=y_1+iy_2$, so that 
\bee
\partial_z =  \frac 1 2 \left(  \frac     {\partial} { \partial  y_1 }  - i\frac  {\partial} {\partial y_2}  \right),\quad 
 \bar{\partial}_z = \frac 1 2 \left( \frac{\partial} { \partial  y_1 }  + i\frac {\partial} {\partial y_2} \right), 
\eee
and introducing the 2-bein  
\bee
\label{2spherebein}
e_1= \frac 2 q dy_1, \quad e_2 = \frac 2 q dy_2,
\eee 
 the metric is  $
ds^2 = e_1^2 + e_2^2$ and 
the dual vector fields are 
\bee
\label{2sphereframe}
E_1=\frac q 2 \frac {\partial}{\partial  y_1}, \quad E_2=\frac q 2 \frac {\partial}{\partial  y_2}.
\eee 
One checks that the two frames are related by a 
 a rotation:
\begin{align}
\label{complexspherical}
E_1 = \cos\alpha \, {\tilde E}_1-\sin\alpha\, {\tilde E_2}, \qquad
E_2 = \sin \alpha\, {\tilde E}_1+\cos\alpha \,  {\tilde E_2}.
\end{align}
This  rotation leads to a gauge change for the associated spin bundles which we will encounter later in our discussion. 

Carrying on with the 2-bein \eqref{2spherebein}, we pick  Clifford generators  in terms of the first two Pauli matrices $\tau_1,\tau_2$:
\bee
 \gamma_1 =  i\tau_1 = \begin{pmatrix}
 0 &  i \\
 i & 0  \end{pmatrix}, 
  \qquad   
  \gamma_2 =  i\tau_2 = \begin{pmatrix}
 0 & 1\\
 -1 & 0  \end{pmatrix}. 
\eee
Computing the spin connection 1-forms from    \eqref{spincon}, 
we find the  non-vanishing component $
\omega_{12} = y_1e_2 -y_2e_1=\frac{2}{q}(y_1dy_2-y_2dy_1)$ and thus the spin  connection \eqref{Gammadef}    as  
\bee
\Gamma =  \frac{i}{q} \tau_3(y_1dy_2-y_2dy_1).
\eee
The  Dirac operator   \eqref{metricdirac} 
is therefore 
\begin{align}
\label{sphe}
\Dslash_{S^2}
  &= \begin{pmatrix} 0 & i(q\partial_z - \frac 1 2    \bar{z})     \\
i(q\bar{\partial}_z - \frac 1 2 z) & 0 \end{pmatrix}.
\end{align}

We now twist this operator with the $n$-th power $H^n$  of the hyperplane bundle, see Appendix~\ref{bundles}, and couple it  to the gauge  potential
 of the Dirac monopole, reviewed in Appendix~\ref{mopoconventions}. Continuing to work  in the patch $U_N$, the gauge potential is 
 \bee
 \label{mopopot}
A_N^n =\frac {n}  {2q} (zd\bar{z} - \bar{z}dz ),
 \eee
so that coupling  amounts to  the   substitutions
\begin{align}
\label{mincouple}
\partial_z  \rightarrow \partial_z - \frac {n} {2q}\bar{z},  \qquad \bar{\partial_z}  \rightarrow \bar{\partial_z} +\frac {n} {2q}z.
\end{align}
We obtain the twisted Dirac operator
\bee
\label{DiracS2}
\Dslash_{S^2,n} = i\begin{pmatrix}
0 & q\partial_z - \frac 1  2(n +1)\bar{z} \\
 q\bar{\partial_z} +\frac 1 2(n -1)z & 0
\end{pmatrix}.
\eee
With the abbreviation
\bee
\label{sndef}
s= \frac 1 2(n -1), \qquad \tilde s = \frac 1 2(n+1),
\eee
we  observe that the operators which appear in the off-diagonal entries here can be written as 
\bee
\label{neat}
q\bar{\partial_z} + sz= q^{-s+1}\bar{\partial_z}q^{s}, \qquad q\partial_z -\tilde s\bar{z} =   q^{\tilde s+1} \partial_z q^{-\tilde s},
\eee
which will be useful later. 
These operators    act on sections of suitable powers of $H$  according to 
\begin{align}
q\bar{\partial_z} + sz&: C^\infty(H^{n-1}) \rightarrow C^\infty(H^{n+1}), \nonumber \\
q\partial_z -\tilde s\bar{z} &: C^\infty (H^{n+1}) \rightarrow C^\infty(H^{n-1}), 
\end{align}
so that  the Dirac operator is a map
\bee
\Dslash_{S^2,n} : C^\infty(H^{n-1}\oplus H^{n+1}) \rightarrow C^\infty(H^{n-1}\oplus H^{n+1})  .
\eee
As reviewed in Appendix~\ref{bundles},  sections of  powers of $H$ can be  described either in terms of local sections $\sec_N:U_N\rightarrow \CC$ and $\sec_S: U_S\rightarrow \CC$ defined on the northern and southern patch respectively and related by a transition function, or in terms of a  function $
\Sec: S^3 \rightarrow \CC $
satisfying an equivariance condition, see \eqref{equiv1} and \eqref{equiv2}. For sections of $H^{n-1}$,  the infinitesimal form of the equivariance condition is
\bee
\label{sconst}
i\lv_3\Sec=s\Sec, 
\eee
while for sections of $H^{n+1}$ it is 
\bee
\label{stildeconst}
i\lv_3\Sec=\tilde{s}\Sec.
\eee

\subsection{The  $\eth$ operator,  $\mathfrak{su}(2)$ generators and an operator for the Chern number} 

In many papers dealing with  the Dirac operator on the 2-sphere,  calculations are carried out in terms of spherical coordinates. In particular, eigenfunctions like the spin spherical harmonics are written as functions of the angles $\beta$ and $\alpha$. In order to facilitate comparisons between our discussion and treatments involving spherical coordinates, we note that in spherical coordinates
\begin{align}
q\bar{\partial_z} + sz& =e^{i\alpha} \left(\partial_\beta +i\frac{1}{\sin \beta } \partial_\alpha
+ s\tan\frac \beta 2  \right), \nonumber \\ q\partial_z -\tilde s\bar{z} &=e^{-i\alpha} \left(\partial_\beta -i\frac{1}{\sin \beta } \partial_\alpha
- \tilde s\tan\frac \beta 2  \right).
\end{align}
It is now easy to establish a link with the "edth" operators which were first  introduced by Penrose and Newman   \cite{NP} and which are frequently used to write the Dirac operator on $S^2$. With
\begin{align}
\label{ethdef}
\eth_s = \partial_\beta +i\frac{1}{\sin \beta } \partial_\alpha
-s\frac{\cos \beta}{\sin \beta },  \qquad 
\bar{\eth}_{\tilde s }= \partial_\beta -i\frac{1}{\sin \beta } \partial_\alpha
+ \tilde  s\frac{\cos\beta }{\sin \beta},
\end{align}
we have the relations
\begin{align}
(q\bar{\partial_z} + sz)e^{is\alpha} = e^{i(s+1)}\eth_s  \quad \text{and} \quad (q\partial_z -\tilde s\bar{z})e^{i\tilde s\alpha} &= e^{i(\tilde s-1)\alpha }\bar{\eth}_{\tilde s }.
\end{align}
They reflect the gauge change from complex to spherical coordinates \eqref{complexspherical}.

In order to relate the discussion here to that of the Dirac operator on Taub-NUT later in this paper we need to understand  how $q\bar{\partial_z} + sz$ and $q\partial_z -\tilde s\bar{z}$ are related to the left-invariant generators $\lv_1,\lv_2, \lv_3$  of the $SU(2)$ right-action on itself, defined in \eqref{lvdef}.
In Appendix \ref{formsfields} we show that  $\lv_\pm = \lv_1\pm i\lv_2$ are raising (+) and lowering (-) operators for the eigenvalue of  $i\lv_3$.  In the description of sections  of powers of $H$ as equivariant functions  with  the differential constraint \eqref{sconst} and \eqref{stildeconst}, the eigenvalue of $i\lv_3$ is related to the power of $H$ according to  \eqref{sndef}. Since $q\bar{\partial_z} + sz$   raises the power of $H$  by two units and 
$q\partial_z -\tilde s\bar{z}$ lowers it by the same amount, we expect the former to be related to $\lv_+$ and the latter to $\lv_-$. This relation was first noticed, using different notation and conventions from ours, in \cite{GMNRS}. We now exhibit it in our notation. 

 Consider a section of $H^{n-1}$ in its equivariant form  \eqref{equidef} as function $\Sec$ of two complex variables $z_1,z_2$  satisfying the constraint \eqref{sconst}.
 We denote  pull-back with the local section $s_N$ \eqref{eNdefined} by $s^*_N$,  so that  in particular
\bee
(s_N^*(X_+\Sec))(z)= i\left(z_1\bar{\partial}_2\Sec - z_2\bar{\partial}_1\Sec\right)\left|_{z_1=\frac {1} {\sqrt{q}} ,z_2=\frac{z} {\sqrt{q}} }\right. .
\eee
 Then  we evaluate 
\bee
i(q\bar{\partial} +sz)(s_N^*\Sec)(z)= i(q\bar{\partial} +sz)\Sec \left( \frac {1} {\sqrt{q}} ,\frac{z} {\sqrt{q}} \right),
\eee
 and  use    the constraint  \eqref{sconst} to  find 
\bee
i(q\bar{\partial} +sz)(s_N^*\Sec)(z)=(s_N^*(\lv_+\Sec))(z).
\eee
Thus, the  operator  $q\bar{\partial} +sz$ acting `downstairs' on a local section  is  the  pull-back of the   $SU(2)$ raising operator $\lv_+$ acting `upstairs' on equivariant functions. 
Similarly, one finds   that  $q\partial - \tilde{s}\bar{z}$ is related to the lowering operator via
\bee
-i(q\partial - \tilde{s}\bar{z})(s_N^*\Sec)(z)=(s_N^*(\lv_-\Sec))(z),
\eee
where we need to use the constraint \eqref{stildeconst}.

Combining these results and introducing the notation 
\bee
 C^\infty (S^3,\CC)_s=\{ F:S^3 \rightarrow \CC\left|  iX_3F= sF \right. \}
\eee
for the space of sections of  $H^{n-1}$ in the equivariant form, we obtain an equivalent operator to 
$\Dslash_{S^2,n}$ acting `upstairs' as 
\begin{align}
\Dslash^*_{S^2,n} = \begin{pmatrix}  0  &   X_-   \\     -X_+   & 0 \end{pmatrix} :
C^\infty (S^3, \CC)_s\oplus C^\infty(S^3,\CC)_{\tilde{s}}
 \rightarrow C^\infty(S^3,\CC)_s\oplus C^\infty (S^3,\CC)_{\tilde{s}},
\end{align}
with $s,\tilde{s}$ defined in \eqref{sndef}.
This operator  commutes with  the operator
\bee
\label{ccoperator}
\hat{n}= 2i\lv_3 +\tau_3: C^\infty (S^3, \CC)_s\oplus C^\infty(S^3,\CC)_{\tilde{s}}
 \rightarrow C^\infty(S^3,\CC)_s\oplus C^\infty (S^3,\CC)_{\tilde{s}}.
\eee
which we  interpret  as `Chern-number operator' since 
it acts as a multiple of the identity with eigenvalue $2s+1 = 2\tilde{s} -1 =n$. We will encounter it in a slightly modified form in our discussion of the Dirac operator on the Taub-NUT space. 
 
 \subsection{Zero-modes on the 2-sphere }

We are now ready to compute the zero modes of  $\Dslash_{S^2,n}$. Working in the patch $U_N$  we 
write the spinor there as  as 
\bee
\psi^N=\begin{pmatrix} \sec_1^N \\ \sec_2^N \end{pmatrix},
\eee
where $\sec^N_1$ is a local section of $H^{n-1}$ and  $\sec^N_2$ a local section of  $H^{n+1}$.
Then 
\begin{align}
\Dslash_{S^2,n}\psi^N=0 \; \Leftrightarrow \; (q\bar{\partial_z} + sz)\sec^N_1 = 0, \qquad  (q\partial_z -\tilde s\bar{z})\sec^N_2=0.
\end{align}
Using the expressions \eqref{neat} we deduce that solutions are of the form  
\bee
\label{norsect}
\sec^N_1(z)=\frac{1}{q^s} p_1(z), \qquad \sec^N_2(z)=q^{\tilde s} p_2(\bar{z}),
\eee
where $p_1$ and $p_2$ are, a priori, two arbitrary holomorphic and,  respectively,  anti-holomorphic functions.
Next,  we implement that they are section of the respective bundles. Using \eqref{patchcond} to switch to the patch $U_S$ 
we require  that 
\bee
\sec^S_1(z)=\frac{1}{q^s} \left(\frac{\bar{z}} {z} \right)^{s}p_1(z)
\eee
is well-defined at $z=\infty$. To check we transform  to $\zeta =1/z$ and find 
\bee
\sec^S_1\left(\frac 1 \zeta\right)=\frac{\zeta^{2s}}{(1+\zeta\bar{\zeta})^s} p_1\left(\frac 1 \zeta\right).
\eee
For this to be well-defined at $\zeta=0$ we require that $p_1$ is a polynomial of degree $\leq 2s = n-1$. In particular, $n$ has to be an integer $\geq 1$ in this case. The dimension of the space of zero modes is $2s+1=n$.

Similarly for the second component, we have to  check if 
\bee
\sec^S_2(z  )=q^{\tilde{s}} \left( \frac{\bar{z}}   {z} \right)^{  \tilde s  } p_2( \bar{z} )
\eee
is well-defined at $z=\infty$. We transform  to $\zeta =1/z$ and find 
\bee
\sec^S_2\left(\frac 1 \zeta \right)=\frac
{  (1+\zeta\bar{\zeta})^{\tilde{s}}    }
{ \bar{\zeta}^ {  2\tilde{s} }   } p_2\left(\frac 1 {\bar{\zeta} }\right),
\eee
which restricts $p_2$ to be a polynomial of degree $\leq -2\tilde s = - n-1$. In particular, $n$ has to be an integer $\leq -1$ in this case. The dimension of the space of zero modes is $-2\tilde{s}+1=-n$.

The zero-modes we have found can be viewed as the pull-back of 
 homogeneous polynomials in two complex variables.  This viewpoint is helpful in understanding the $SU(2)$ action on the zero-modes, and also provides a link with the zero-modes on the Taub-NUT space in the next section.  Pulling back
\bee
\label{holpol}
P_1(z_1,z_2) =\sum_{k=0}^{n-1} a_k z_1^{n-1-k}z_2^{k}, \qquad n\geq 1,
\eee
with  the local section $s_N:U_N\rightarrow S^3$  \eqref{eNdefined}
gives all  the zero modes in the  case  $n>0$.  Indeed, 
\bee
(s^*_N P_1)(z) =P_1\left(\frac {1} {\sqrt{q}}, \frac {z}{\sqrt{ q}}\right)=\frac{1}{q^{s}}\sum_{k=0}^{n-1} a_k z^k, \qquad n\geq 1
\eee
 is the general form of $\sec_1^N$.
When $n<0$, we start with a homogeneous and anti-holomorphic polynomial 
\bee
\label{antiholpol}
P_2(\bar{z}_1,\bar{z}_2) =\sum_{k=0}^{-n-1} a_k \bar{z}_1^{-n-1-k}\bar{z}_2^{k}\qquad n\leq 1.
\eee
Again we  pull-back with $s_N$ to obtain
\bee
(s^*_NP_2)(\bar{z})=P_2\left(\frac {1} {\sqrt{q}}, \frac {\bar{z}}{\sqrt{ q}}\right)=q^{\tilde{s}} \sum_{k=0}^{-n-1} a_k \bar{z}^k,
\qquad n\leq 1,
\eee
which is the general form of $\sec_2^N$.

Summing up, the zero modes of $\Dslash_{S^2,n}$ take the following form on $U_N$:
\bee
\label{S2zeromodes}
\psi^N(z) =  \begin{pmatrix}  q^{\frac 12 (1-n)}\sum_{k=0}^{n-1} a_k z^k \\ 0 \end{pmatrix} \; \text{if} \; n\geq 1, \quad \psi^N(\bar{z})=  \begin{pmatrix}  0 \\ q^{\frac 12 (1+n)} \sum_{k=0}^{-n-1} a_k \bar{z}^k  \end{pmatrix} \;\text{if} \; n\leq -1.
\eee  
\subsection{Zero-modes as irreducible  $SU(2)$ representations}

The $|n|$-dimensional space of zero modes of $\Dslash_{S^2,n}$ is naturally acted on by the double cover  $SU(2)$ of the isometry group of the 2-sphere. The quickest way to see that the space of zero modes is actually the $|n|$-dimensional irreducible representation of $SU(2)$ is to use the description of the zero modes as homogeneous polynomials in the two complex variables $z_1,z_2$  in \eqref{holpol} and \eqref{antiholpol}. As reviewed in Appendix~\ref{harmonic}  before equations \eqref{holobasis} and \eqref{antiholobasis},  polynomials of the forms \eqref{holpol}  and   \eqref{antiholpol}  span  the irreducible  $SU(2)$ representations of dimension $n$ for $n>0$ and $-n$ for  $n<0$.

Explicitly, an $SU(2)$  element  \bee
U=\begin{pmatrix}  \phantom{-}b  & \bar{a} \\ -a & \bar{b} \end{pmatrix}, \qquad |a|^2+|b|^2=1,  
\eee
acts on the polynomials \eqref{holpol} and \eqref{antiholpol} via pull-back with the inverse
\bee
U^{-1}=\begin{pmatrix} \bar{ b}  & -\bar{a} \\ a & \phantom{-}b \end{pmatrix}, 
\eee
i.e., by mapping the arguments $(z_1,z_2)$ according to
\bee
\label{su2}
\begin{pmatrix} z_1 \\ z_2 \end{pmatrix} \mapsto \begin{pmatrix} \bar{ b}  & -\bar{a} \\ a & \phantom{-}b \end{pmatrix}
\begin{pmatrix} z_1 \\ z_2 \end{pmatrix} = \begin{pmatrix} \bar{b} z_1 -\bar{a}z_2 \\ az_1 + bz_2 \end{pmatrix},
\eee
and $(\bar{z}_1,\bar{z}_2)$ correspondingly.

The transformation of the zero-modes\eqref{S2zeromodes} under the $SU(2)$ action
is induced by pulling back  the action \eqref{su2}.  The non-trivial nature of the line bundles implies an additional phase factor or multiplier, as we shall now show. We introduce the notation $u^{-1}$
for the mapping induced by \eqref{su2}  on the quotient $z=z_2/z_1$:
\bee
 u^{-1}:z\mapsto \frac{a+bz}{\bar{b} -\bar{a}z}.
\eee
Exploiting $|a|^2+|b|^2=1$, the function $q$ \eqref{qdef} satisfies
\bee
\label{qtrans}
q(u^{-1}(z))=\frac{q(z)}{(\bar{b} -\bar{a}z) (b-a\bar{z})}.
\eee

For any  local section  $\sec: U_N\rightarrow \CC$ which is the pull-back of a function $\Sec:S^3\rightarrow \CC$ satisfying the equivariance condition \eqref{equiv2}, we define 
\bee
\label{suupstairs}
\rho_s(U)\sec =s^*_N (\Sec\circ U^{-1}).
\eee
Using \eqref{equiv2} and \eqref{qtrans}, one checks that 
\bee
\label{suact}
(\rho_s(U)\sec)(z) =\mu_s(U;z)\sec(u^{-1}(z)),
\eee
where the multiplier $\mu_s$ is 
\bee
\mu_s(U;z) = \left(\frac{\bar{b}-\bar{a}z}{b-a\bar{z}}\right)^s.
\eee
It satisfies
\bee
\mu_s(U_1;z)\mu_s(U_2;U_1^{-1}z) = \mu_s(U_1U_2,z),
\eee
which ensures that \eqref{suact} is  an action. 

For $\sec(z) =q^{-s}p(z)$, where $p$ is a polynomial of degree $\leq 2s$, we note
\bee
(\rho_s(U)\sec)(z,\bar{z}) = \frac{1}{q^s} (\bar{b}-\bar{a}z)^{2s} p\left(\frac{a+bz}{\bar{b} -\bar{a}z}\right).
\eee
Since $p$ has degree $\leq 2s$, this is again a product of $q^{-s}$ with a polynomial of degree $\leq 2s$.

We conclude that the local sections of the form $\sec^N_1$ in  \eqref{norsect}  form the irreducible representation of $SU(2)$
of dimension $n=2s+1$ and spin $j=s$. A similar argument shows that, for $n<0$,  the local sections $\sec^N_2$ in  \eqref{norsect} form an irreducible representation of dimensions $-n=-2\tilde{s} +1$ and spin   $j=-\tilde{s}$.

\subsection{Zero-modes on $\RR^3$}
In this section we show that the zero-modes of the Dirac operator $\Dslash_{S^2,n}$ give rise to zero-modes of a certain massive Dirac operator on Euclidean 3-space. This will provide valuable intuition for analysing the zero-modes on the Taub-NUT manifold in the next section.

The standard Dirac operator on $\RR^3$ associated to  the  flat metric in Cartesian coordinates
$
 ds^2 =d x_1^2 +dx_2^2 +dx_3^2
 $ is simply
 \bee
\Dslash_{\RR^3} = i\tau_j\partial_j. 
 \eee
 However, the Cartesian form is not convenient in the current context, for two reasons. The action of rotations on spinors is more complicated in the Cartesian frame since it is not rotationally invariant. Furthermore, the monopole
 gauge potential takes its simplest form in coordinates adapted to the foliation of $\RR^3$ into spheres. 

Using again the complex coordinate $z$ on the sphere without the  South Pole,  we   write the  flat metric of $\RR^3$  as 
\bee
\label{rzcoords}
ds^2 =dr^2 + \frac{4r^2} {q^2}  dzd\bar{z},
\eee
and obtain a  3-bein by adding $dr$ to the rescaled 2-bein \eqref{2spherebein}:
\bee
e_1= \frac {2r }{q} dy_1, \quad e_2 = \frac {2r}{ q} dy_2, \quad e_3=dr.
\eee 
The spin connection forms are  
\bee
\label{spincon3}
\omega_{12}=\frac 2 q (y_1dy_2-y_2dy_1), \quad \omega_{23}=\frac{2}{q}dy_2, \quad \omega_{13}=\frac 2 q dy_1,
\eee
and the spin connection is
\begin{align}
\Gamma^{(3)} = \frac i 2 (\omega_{12}\tau_3 +\omega_{23}\tau_1 +\omega_{31}\tau_2)=\frac i q ((y_1dy_2-y_2dy_1)\tau_3 +dy_2\tau_1-dy_1\tau_2)
\end{align} 
With the  dual vector fields  
\bee
E_1=\frac {q} {2r} \frac {\partial}{\partial  y_1}, \quad E_2=\frac {q} {2r} \frac {\partial}{\partial  y_2},
\quad E_3= \partial_r,
\eee 
and  the gamma matrices $\gamma_j=i\tau_j$, $j=1,2,3$, 
the  Dirac operator on $\RR^3$ coupled to the monopole gauge field  \eqref{mopopot} is 
\begin{align}
\Dslash_{\RR^3,n} &=\sum_{j=1}^3\gamma_j \iota_{E_j} (d+ A^n_N +\Gamma^{(3)})
\nonumber \\
&= i\begin{pmatrix} \partial_r+\frac 1r& 0 \\ 0 & -\partial_r-\frac 1 r \end{pmatrix} + 
\frac 1 r \Dslash_{S^2,n} ,
 \end{align}
 where 
 $\Dslash_{S^2,n}$ is defined in \eqref{DiracS2}.
$ \Dslash_{\RR^3,0}$ is related to $\Dslash_{\RR^3}$ by a gauge transformation.
  
  We will discuss the zero modes of $\Dslash_{\RR^3,n}$ in the context of a deformed version of this operator,
   where the deformation parameter is an inverse length or   mass (in units where $\hbar=c=1$). The operator we consider may be thought of as a   singular limit of the Dirac operator coupled to a smooth non-abelian BPS monopole \cite{JR}. Callias  proved  an index theorem  for smooth non-abelian  BPS monopoles in \cite{Callias} and considered  singular limit where the Higgs field is taken to be constant  in \cite{Callias2}. This is the limit we consider here.  A different singular limit, first considered in \cite{Nahm}, requires the Higgs field to satisfy the abelian Bogomol'nyi equation, see also  \cite{CF} for a recent discussion of the associated Dirac equation and plots of its zero-modes. 
   
   We obtain  our  operator via  dimensional reduction of a Dirac operator in $\RR^4$ coupled to a Dirac monopole in $\RR^3$ and a constant connection   $\frac i \Lambda dx_4$, where $\Lambda$ is a non-negative length scale and $x_4$ a coordinate for the auxiliary fourth dimension. Working again with the coordinates $r,z$  used in \eqref{rzcoords}, the metric on $\RR^4$ is
  \bee
  ds^2=dr^2 +\frac{4r^2} {q^2}  dzd\bar{z} +dx_4^2.
  \eee
    With the Euclidean Dirac matrices
  \bee
\label{gamma4}
 \gamma_i =
\bpm 0 & \tau_j\cr -\tau_j& 0 \epm, \; j=1,2,3\qquad \gamma_4= \bpm 0 & -i\bone \cr -i\bone & 0 \epm ,
\eee
we  have the commutators
\bee
\label{gammacom}
[\gamma_4,\gamma_i] =  2i\bpm \tau_i & \phantom{-} 0\cr 0 &-\tau_i  \epm\quad
\mbox{and} \quad [\gamma_i,\gamma_j] =-2i\epsilon_{ijk}
\bpm\tau_k & 0 \cr 0 & \tau_k \epm.
\eee
 Noting that the  non-vanishing  connection 1-forms are as in \eqref{spincon3}, the spin connection  is  $4\times 4$ matrix which can be written in terms of the spin connection $\Gamma^{(3)}$ as 
  \bee
  \Gamma^{(4)}= \begin{pmatrix}  \Gamma^{(3)}& 0  \\ 0&  \Gamma^{(3)} \end{pmatrix}.
  \eee
 With a $U(1)$ gauge potential which combines the Dirac monopole \eqref{mopopot} with a constant component in the $x_4$-direction, 
  \bee
 A= \frac {n}  {2q} (zd\bar{z} - \bar{z}dz )+ \frac i \Lambda dx_4,
  \eee
  the twisted Dirac operator has the general form \eqref{gendirac}.  For spinors which do not depend on the auxiliary coordinate $x_4$,  it simplifies to 
  \begin{align}
  \label{Dirac4}
  \Dslash_{\Lambda,n} &= \sum_{\alpha=1}^3\gamma_j \iota_{E_j} (d+ A^n_N +\Gamma^{(4)})
  +\frac i \Lambda \gamma_4 \nonumber \\
  &=
\begin{pmatrix}  0 &   -i\Dslash_{\RR^3,n} +\frac 1 \Lambda \, \bone   \\    i\Dslash_{\RR^3,n} +\frac 1 \Lambda \, \bone  & 0 \end{pmatrix}.
\end{align}
It is easy to check that the zero-modes \eqref{S2zeromodes} of $\Dslash_{S^2,n}$ give rise to the following square-integrable zero-modes of \eqref{Dirac4} on the open set $\RR^+\times U_N$:
\bee
\label{massivesolutions}
\Psi^N =   \frac{e^{-\frac r \Lambda }}{r} \begin{pmatrix} 0 \\ 0 \\ q^{\frac 12 (1-n)}\sum_{k=0}^{n-1} a_k z^k \\ 0 \end{pmatrix} \; \text{if}\; \; n\geq 1, \quad \Psi^N= \frac{e^{-\frac r \Lambda }}{r}  \begin{pmatrix}  0 \\ q^{\frac 12 (1+n)} \sum_{k=0}^{n-1} a_k \bar{z}^k  \\0 \\0 \end{pmatrix} \;\text{if} \;\; n\leq -1.
\eee  
These solutions are singular at $r=0$ but square integrable on $\RR^3$. When we take the limit  $\Lambda=\infty$ we lose the square-integrability.  Similarly, allowing for spinors on the 2-sphere which are not zero-modes of $\Dslash_{S^2,n}$ generates solutions which diverge at $r=0$ faster than $1/r$. Such solutions are also not square-integrable. 

We have exhibited an $|n|$-dimensional space of normalisable zero-modes  of the deformed or `massive' Dirac operator \eqref{Dirac4}. In the context of this paper we are interested in these zero-modes because they provide valuable intuition for understanding the normalisable zero-modes of  the twisted Dirac operator on the Taub-NUT manifold in the next section. We do not claim to have proved that all normalisable zero modes are of the form \eqref{massivesolutions} although we expect this to be the case. A rigorous  discussion would need to address issues of self-adjointness, see \cite{Callias2} for the case of $n=1$ and \cite{ES} for a recent and  general treatment of zero-modes of magnetic Dirac operators on $\RR^3$.

\section{Twisted Dirac operators on the Taub-NUT manifold}

\subsection{Dirac operators on self-dual 4-manifolds with rotational symmetry}
\label{gensect}
Although we are primarily interested in the Taub-NUT manifold in this paper, we initially work in a more general framework and  give the form of the Dirac operator
for  four-manifolds with  isometry group $SU(2)$  or $SO(3)$, acting   with  generically 3-dimensional orbits,  and a  self-dual Riemann tensor. A partial list of  examples  of such `gravitational instantons' can  be found in \cite{EGH}.   In particular, we have in mind the Atiyah-Hitchin manifold which was  considered in \cite{AMS} alongside the Taub-NUT manifold  as a candidate for a geometric model of matter.  The metrics can be parametrised in terms of suitable  $SU(2)$ or $SO(3)$ orbit parameters (e.g. our Euler angles or complex coordinates) and a transverse, radial coordinate $r$.
In terms of the left-invariant 1-forms $\sigma_j$, $j=1,2,3$,  and radial functions $f,a,b,c$, the metrics take the form
\bee
\label{athi}
 ds^2 = f^2dr^2 + a^2\lf_1^2 +b^2\lf_2^2 +c^2\lf_3^2.
\eee
The function $f$ 
may be chosen freely, different choices corresponding to 
different definitions of the radial coordinate $r$.
We  introduce  the tetrad
\bee
\label{tetrad}
e_1= a \lf_1, \quad e_2= b \lf_2, \quad e_3=c\lf_3, \quad e_4 =-f dr.
\eee
We use the orientation discussed in  \cite{AMS}. Since  the left-invariant 1-forms $\lf_i$, $i=1,2,3$,  have the opposite sign of the left-invariant 1-forms used in \cite{AMS} (see also Appendix~\ref{parametrisations})  the resulting volume element is
\begin{align}
\label{volume}
dV =  e_1\wedge e_2\wedge e_3\wedge e_4  =fabc  \, dr \wedge \lf_1 \wedge \lf_2\wedge \lf_3 = fabc\sin \beta d r \wedge d \beta \wedge d\alpha \wedge d \gamma. 
\end{align}

The self-duality of the Riemann tensor  with respect to the orientation
implies
\bee
\label{dual}
 \frac {2bc} {f}\frac { da}{dr} = (b-c)^2 -a^2,\; \; \text{+ cycl.},
\eee
where `+ cycl.' means we add the two further equations
obtained by cyclic permutation of $a,b,c$.  
Solving \eqref{spincon} for the spin connection,   we find
\begin{align}
\omega_{14} &= (1-A)\lf_1, \quad \omega_{24} = (1-B)\lf_2, \quad \omega_{34} = (1-C)\lf_3, 
\nonumber \\
\omega_{23}& =-A\lf_1, \;\;\;\quad \quad  \omega_{31}= -B\lf_2, \;\;\;\quad \quad \omega_{12} = -C\lf_3,
\end{align}
where 
\bee
A= \frac{b^2+c^2 -a^2}{  2 bc}, \quad B=\frac{a^2 + c^2 -b^2}{ 2ac}, 
\quad C=\frac {a^2 +b^2 -c^2 }{ 2ab}.
\eee

The vector fields dual to the tetrad \eqref{tetrad} are 
\bee
\label{dualtetrad}
E_1 = \frac {1} {a} \lv_1, \quad
E_2 =\frac{1} {b} \lv_2, \quad
E_3 = \frac{1} {c} \lv_3,\quad
E_4 =-\frac {1} {f} \frac {\partial} {\partial r}, 
\eee
where $\lv_1,\lv_2$ and $\lv_3$  are the left-invariant vector fields  on $SU(2)$ \eqref{lvectors}.  For our purposes, the advantage of  working with the frames \eqref{tetrad} and \eqref{dualtetrad} is that they are rotationally invariant. This results in a choice of gauge for the Dirac operator and the bundle of spinors where the $SU(2)$ action is particularly simple. Note that many treatments of the Dirac operator on the Taub-NUT manifold (e.g., in \cite{CH}) use a different gauge.

For many calculations it is convenient to use a  proper radial distance coordinate $R$  defined via 
\bee
dR =fdr,
\eee
and we frequently  do this in the remainder of this section. We are interested in the general form of Dirac operators on metrics like \eqref{athi} and  coupled to a spherically symmetric, abelian ($U(1)$ or $\RR$) connection with self-dual curvature. Locally, the gauge potential for such a connection can be written in terms of the left-invarian 1-forms  as 
\bee
\label{sdcon}
\gp = A_1\lf_1+  A_2\lf_2+ A_3\lf_3,
\eee
where $A_1,A_2$ and $A_3$ are functions of $R$ only.  The curvature  is
\bee
 \gf= d\gp= \frac{1}{a}\frac{dA_1}{dR}e_1\wedge e_4 -\frac{A_1}{bc}e_2\wedge e_3+
\frac{1}{b}\frac{dA_2}{dR}e_2\wedge e_4 -\frac{A_2}{ca}e_3\wedge e_1+
\frac{1}{c}\frac{dA_3}{dR}e_3\wedge e_4 -\frac{A_3}{ab}e_1\wedge e_2, 
\eee
which is self-dual if 
\bee
\label{adual}
\frac{dA_1}{dR}=-\frac{a}{bc}A_1,\quad \frac{dA_2}{dR}=-\frac{b}{ac}A_2, \quad\text{and} \quad   \frac{dA_3}{dR}=-\frac{c}{ab}A_3.
\eee
In the following we write $D_j=\lv_j+iA_j$, $j=1,2,3$,  for the associated covariant derivatives.

Working again with the Euclidean  $\gamma$-matrices \eqref{gamma4} and associated commutators \eqref{gammacom}, the Dirac operator \eqref{gendirac} associated to the metric \eqref{athi} and the connection \eqref{sdcon}  takes the form
\bee
\label{SDdirac}
\Dslash_\gp = \bpm 0 & T_\gp^{\dagger} \cr T_\gp & 0 \epm ,
\eee
where
\begin{align}
\label{Tgen}
T_\gp^{\dagger} 
&=  \frac i  f  \frac {\partial }{\partial r} -\frac i 2 \left(\frac 1 a +  \frac 1 b + \frac 1 c \right)
+\frac  1  a\tau_1 D_1  +\frac 1 b \tau_2 D_2  +\frac   1 c \tau_3D_3, \quad \nonumber \\
T_\gp &=  \frac i  f  \frac {\partial }{\partial r} + i \left(\frac {A}{a} +  \frac {B} { b} +  \frac {C} {c}\right) -\frac  i 2 \left(\frac 1 a + \frac 1 b + \frac 1 c \right)
-\frac  1  a \tau_1 D_1  -\frac 1 b \tau_2 D_2  -\frac   1 c \tau_3D_3.
\end{align}

As a result of the rotational (left-)invariance of the metric, the tetrad \eqref{tetrad} and the connection \eqref{sdcon}, the Dirac operator commutes with the vector fields $\rv_1,\rv_2$ and $\rv_3$ \eqref{leftgenerators} generating the left-action of $SU(2)$ or $SO(3)$ on the manifold. This is easily checked explicitly, since the left-generators commute with the right-generators $\lv_1,\lv_2$ and $ \lv_3$ and any function of the radial coordinate $r$,  see   Appendix~\ref{formsfields} for further details.  The operators $i\rv_j, j=1,2,3$, play the role of the total angular momentum operators, combining both  orbital and spin contributions. In our rotationally symmetric gauge, the total angular momentum operators only act on the argument of the spinors and do not mix their components. 

To check that  $T_\gp$ and $T_\gp^\dagger$ are actually each others' adjoints with respect to the $L^2$  inner product based on the  volume element \eqref{volume} we
note that, as a consequence of the self-duality equations \eqref{dual}, 
\bee
\label{useful}
\frac {1}{ abcf }\frac {\partial} { \partial r} abc
=\frac {A-1}{a} + \frac {B-1} { b} + \frac {C-1} {c} + \frac {1}{ f}\frac  {\partial}
{ \partial r}.
\eee

To end this section we show that, for non-compact self-dual 4-manifolds,    $T_\gp^{\dagger}$ has a trivial kernel. This is a special case of a vanishing theorem for Dirac operators on non-compact self-dual manifolds coupled to line bundles with self-dual connections  proved in \cite{SSZ}. However, the following short  proof  for the spherically symmetric case contains some illuminating details. In particular, we see  an interesting relation to the Dirac operator on the squashed 3-sphere. 

The Dirac operator on the 3-sphere with metric 
\bee
ds^2 =a^2\sigma_1^2 + b^2\sigma_2^2 + c^2 \sigma_3^2
\eee
at a fixed value of $r$ (or, equivalently, for real constants $a,b$ and $c$)
and coupled to the connection \eqref{sdcon} at fixed value of $r$  is 
\begin{align}
\label{Diracs3}
\Dslash_{S^3,\gp} &= \frac{i}{a} \tau_1 D_1 +\frac{i}{b} \tau_2 D_2+ \frac{i}{c} \tau_3 D_3 +\frac 1 2 \left(\frac A  a +\frac B b + \frac Cc\right).
\end{align}
Therefore 
we can write
\begin{align}
T^{\dagger}_\gp 
&=  \frac i  f  \frac {\partial }{\partial r} -i\Dslash_{S^3,\gp} +\frac i 2 \left( \frac {A-1}{a} + \frac {B-1} { b} + \frac {C-1} {c}  \right), \nonumber \\
T_\gp &=  \frac {i}{f}\frac  {\partial}{\partial r}+i\Dslash_{S^3,\gp} +\frac i 2  \left( \frac {A-1}{a} + \frac {B-1} { b} + \frac {C-1} {c}  \right).
\end{align}

We can simplify these expressions by introducing  the differentiable function $
\nu =\sqrt{|abc|}$,  noting that, for Riemannian metrics, the  functions $a,b$ and $c$ solving \eqref{dual} cannot pass through zero and therefore do not change sign. Then, using \eqref{useful},
we  obtain the symmetric formulae
\bee
\label{tns3}
T_\gp=\frac i \nu  \frac{\partial}{\partial R} \, \nu   +i\Dslash_{S^3,\gp} , \qquad 
T^\dagger_\gp=\frac i \nu \frac{\partial}{\partial R} \, \nu   -i\Dslash_{S^3,\gp},
\eee
and therefore
\bee
T_{\gp}T_\gp^{\dagger} = -\left( \frac{1}{\nu}\frac{\partial}{\partial R}\nu\right)^2 + \Dslash_{S^3,\gp}^2 + \frac{\partial\Dslash_{S^3,\gp}}{\partial R}.
\eee

 Using the self-duality equations \eqref{dual} and  \eqref{adual} as well as the  commutation relations $[\lv_i,\lv_j]=\epsilon_{ijk}\lv_k$, one finds after a lengthy computation
\begin{align}
T_{\gp} T_\gp^{\dagger} &= -\left(\frac{1}{\nu}\frac{\partial}{\partial R}\nu\right)^2 
- \frac{D_1^2}{a^2} -\frac{D_2^2}{b^2} -\frac{D_3^2 }{c^2} 
+\frac{i}{a^2}\tau_1 D_1 + \frac{i}{b^2}\tau_2D_2 + \frac{i}{c^2}\tau_3D_3 
 \nonumber \\
&+ \left(\frac{a^2+b^2+c^2}{4abc}\right)^2 + \frac{d}{dR}\left(\frac{a^2+b^2+c^2}{4abc}\right).
\end{align}
Now we observe that 
\bee
 \frac{1}{abc}\partial_R abc \partial_R= \left( \frac{1}{\nu}\frac{\partial}{\partial R}\nu\right)^2 -\frac{1}{\nu}\frac{d^2 \nu}{dR^2},
\eee
and complete the square to obtain
\begin{align}
\label{vanishing}
T_{\gp}T_{\gp}^{\dagger} &= -\frac{1}{abc}\partial_R abc \partial_R - 
\frac{1}{a^2} \left(D_1 - \frac{i}{2}\tau_1\right)^2 - 
\frac{1}{b^2}\left(D_2 - \frac{i}{2}\tau_2\right)^2
 - \frac{1}{c^2}\left(D_3 - \frac{i}{2}\tau_3\right)^2 
 + W,
\end{align}
with 
\bee
W =-\frac{1}{\nu}\frac{d^2 \nu}{dR^2} - \frac{1}{4a^2} - \frac{1}{4b^2} - \frac{1}{4c^2} + \left(\frac{a^2+b^2+c^2}{4abc}\right)^2 + \frac{d}{dR}\left(\frac{a^2+b^2+c^2}{4abc}\right).
\eee
However, this function vanishes identically as a consequence of the self-duality equations \eqref{dual}. 

Taking the expectation value of the  identity \eqref{vanishing} and integrating by parts, one deduces that  any zero-mode  of  $T_{\gp}^{\dagger}$  would have to be covariantly constant. On a non-compact manifold this is impossible for a normalisable spinor. Therefore $T^\dagger_\gp$ cannot have any zero-modes.

\subsection{Dirac operators on Taub-NUT coupled to self-dual  $\RR$-gauge fields}

We now insert  the solution of the self-duality equations \eqref{dual} which gives rise to the Taub-NUT metric:
\bee
\label{TNprofile}
a=b=r\sqrt{V}, \quad c=\frac{L}  { \sqrt{V} }, \quad f=-\frac{b}{r}=-\sqrt{V}, 
\eee
where 
\bee
V=1+\frac L r,
\eee
and $L$ a positive parameter, which plays the role of a length scale in the current context.  
Substituting into \eqref{Tgen}, we have 
\begin{align}
T^\dagger& = \frac{i}{\sqrt{V}}\left(-\partial_r  -\frac 1 r -\frac{V}{L} \left(i\tau_3X_3 +\frac 1 2 \right) + \frac 1 r (-i\tau_1X_1 -i\tau_2 X_2)\right), \nonumber \\
T & = \frac{i}{\sqrt{V}}\left(-\partial_r  -\frac 1 r +\frac{V}{L}\left(i\tau_2X_3 +\frac 1 2 \right) + \frac{L}{2r^2V}+ \frac 1 r \left(i\tau_1X_1 +i\tau_2X_2\right)\right).
\end{align}

The Dirac operator on the Taub-NUT manifold has been studied extensively in the literature, starting with \cite{EI,KS,BB}. It 
 does not have normalisable zero-modes. However, zero-modes appear  when the Taub-NUT Dirac operator  is twisted by an  abelian connection with a self-dual curvature, i.e.,  with a special solution of the Maxwell equations. This connection was first noted and coupled to the Dirac operator  by Pope  in \cite{Pope1}. Its curvature  turns out to have a finite $L^2$-norm, and has played a role as a BPS state in tests of S-duality \cite{Weinberg,Gauntlett}. 

One way to understand the origin of this solution in the Taub-NUT geometry is to note that 
the self-duality equations \eqref{dual} for the coefficient functions in the TN case  ($a=b$) include the equation
\bee
2\frac{dc}{dr}=-\frac{fc^2}{ab},
\eee
which, together with \eqref{adual},  implies that 
\bee
\label{our}
\gp = K c^2 \lf_3
\eee
 has a self-dual exterior derivative, for any constant $K$:
\bee
\gf = d\gp=K\frac{c^2}{ab}(e_4\wedge e_3 + e_2 \wedge e_1) =K(\frac{c^3}{ar} dr \wedge\lf_3 + c^2 \lf_2\wedge \lf_1),
\eee
where we used $f=-b/r$ and $e_4=-fdr$.
Since $\gf$ is exact, it is automatically closed. By self-duality it is  co-closed and harmonic. 

There is no natural normalisation of $\gf$. In particular, since the Taub-NUT manifold is diffeomorphic to $\RR^4$, there are no non-trivial 2-cycles and we cannot normalise $\gf$ by  its flux.  We would like to interpret $\gf$ as the curvature of a connection, but, as explained in our Introduction,   in the absence of non-trivial  2-cycles we allow the gauge group to be $(\RR,+)$ rather than $U(1)$.  Nonetheless we will adopt a convenient normalisation, namely we  pick $K$ so that $\gp$ can be interpreted as a   connection form on $S^3$  (viewed as the total space of the Hopf bundle)  for large $r$. With $K=i/(2L^2)$, we have 
\bee
\label{ouragain}
\gp = i\frac{c^2}{2 L^2} \lf_3= \frac{i}{2} \frac{r}{r+L}\lf_3.
\eee
Taking the limit $r\rightarrow \infty$ we obtain the form $\frac  i 2 \lf_3$, 
which, in analogy with \eqref{conn}, can be interpreted as a connection 1-form on $S^3$.

The real 2-form
\bee
\omega:= - \frac{i \gf}{2\pi}=\frac{1}{4\pi }\left( \frac{r}{r+L}\lf_2\wedge \lf_1 +  \frac{L} {(r+L)^2} dr \wedge \lf_3\right)
\eee
was tentatively interpreted as  the electric field in a geometric model of the  electron in \cite{AMS},  where the roles of electric and magnetic fields were swapped relative to the discussion here. 
 In that context, the normalisation $  \int_{\text{TN}}\omega \wedge \omega     = 1 $
was related to the electron charge being $-1$.

Minimally coupling the connection \eqref{ouragain} to the Dirac operator, and allowing for  spinors with charge $p\in \RR$, we obtain the 
operator
\bee
\label{TNdirac}
\Dslash_p = \bpm 0 & T_p^{\dagger} \cr T_p & 0 \epm ,
\eee
where
\begin{align}
T_p^\dagger
&= \frac i  f  \frac {\partial }{\partial r} -\frac i 2 \left(\frac 1 a +  \frac 1 b + \frac 1 c \right)
+\frac  1  a\tau_1 \lv_1  +\frac 1 b \tau_2\lv_2  +\frac   1 c \tau_3\left(\lv_3+\frac {ipc^2} {2L^2} \right)  \nonumber \\
& = \frac{i}{\sqrt{V}}\left(-\partial_r  -\frac 1 r -   \frac{V}{2L} +\tau_3 \left(\frac{p}{2L} - \frac{iV}{L} X_3\right) -\frac i r (\tau_1X_1 +\tau_2X_2)\right), \nonumber \\
T_p 
& =  \frac i  f  \frac {\partial }{\partial r} + i \left(\frac {A}{a} +  \frac {B} { b} +  \frac {C} {c}\right) -\frac  i 2 \left(\frac 1 a + \frac 1 b + \frac 1 c \right)
-\frac  1  a \lv_1\tau_1  -\frac 1 b \lv_2\tau_2  -\frac   1 c\tau_3  \left(\lv_3 +\frac {ipc^2} {2L^2} \right) \nonumber \\
& = \frac{i}{\sqrt{V}}\left(-\partial_r  -\frac 1 r +\frac{V}{2L} + \frac{L}{2r^2V}
+\tau_3 \left(\frac{iV}{L}X_3-\frac{p}{2L}\right)+ \frac i r (\tau_1 X_1 +\tau_2X_2)\right).
\end{align}
Like the Dirac operator \eqref{SDdirac},  the Dirac operator \eqref{TNdirac} commutes with the generators $\rv_1,\rv_2$ and $ \rv_3$  of the  $SU(2)$ left-action. The equality $a=b$ for the Taub-NUT metric further implies that \eqref{TNdirac} also commutes with the right-generator
\bee
\label{hat3}
\hat{\lv}_3= 
\lv_3-\frac i 2  \begin{pmatrix} \tau_3 &  0 \\
0 & \tau_3 \end{pmatrix}.
\eee
This follows form the identity $[\lv_3-\frac i 2 \tau_3, (\lv_1\tau_1+\lv_2\tau_2)]=0$. The operator $\hat{\lv}_3$ is the lift of the generator $\lv_3$ of the central $U(1)$ inside the isometry group $U(2)$ to spinors.

\subsection{ Zero-modes  and $SU(2)$ representations}
In order to write down the zero modes of \eqref{TNdirac} explicitly, 
we introduce the dimensionless radial coordinate  $\rho = r/L$, so that 
 $V = 1 + 1/\rho$.
Further using the notation $X_\pm =X_1\pm iX_2$  of  Appendix \ref{formsfields} we 
have 
\begin{align}
T_p^\dagger &= \frac {i} { L \sqrt{V}   }    \begin{pmatrix}
-\partial_\rho - \frac {1}{ \rho}   - \frac {V}{ 2} - iV\lv_3  + \frac{p} {2} &
- \frac {i} { \rho} \lv_- \\
-\frac {i}{\rho}\lv_+ &
-\partial_\rho - \frac  {1}{ \rho} - \frac V  2 + iV\lv_3  -\frac{p}{2}
\end{pmatrix}, \nonumber \\
T_p &= \frac {i}{ L\sqrt{V}}\begin{pmatrix}
-\partial_\rho - \frac {1}{ \rho}  +\frac V 2 + \frac{1}{ 2\rho^2V} +  iV\lv_3 - \frac{p}{2} &
\frac {i}{ \rho}\lv_- \\
\frac {i}{ \rho}\lv_+&
-\partial_\rho - \frac {1}{ \rho} +\frac V 2+ \frac {1}{2\rho^2V}  -iV\lv_3 + \frac p 2 
\end{pmatrix}.
\end{align}

We are now ready  to solve
\bee
\label{zeroeq}
\Dslash_p\Psi=0
\eee
 for a 4-component spinor $\Psi$ and interpret Pope's formula \eqref{popedim} for the dimension of the space of solutions. We will  exhibit  the zero-modes in our complex notation and decompose them under the action of $SU(2)$. It follows from our general discussion in Sect.~\ref{gensect} that  the operator $T_p^\dagger$ has no zero modes.  We   therefore only  need to  consider the top two components of $\Psi$. 

 The operator  $T_p$
commutes with the generators $\rv_1,\rv_2$ and $\rv_3$  of the $SU(2)$ left-action and  the lifted right-generator $\hat{\lv}_3$ \eqref{hat3}.  We can therefore  assume eigenspinors to be eigenstates of  $\rv_3, \hat{\lv}_3$ and the  (scalar) Laplace operator on the round 3-sphere $\Delta_{S_3}$,  see \eqref{s3laplace} for an expression in terms of both left- and right-generators of the $SU(2)$ action. These three operators mutually commute, and  common eigenfunctions are discussed in Appendix~\ref{harmonic}. With the eigenvalues of $\Delta_{S_3}$ being $-j(j+1)$ for $j=0,\frac 1 2,1,\frac 32 \ldots$, the eigenvalues  $m$ of  $\rv_3$ and $s$ of  $\lv_3$  both lie in  the range $-j,-j+1,\ldots,j-1,j$. As explained in the appendix, eigenfunctions can be expressed as homogeneous polynomials in $z_1,z_2,\bar{z}_1,\bar{z}_2$, with holomorphic polynomials for the case $s=j$ and anti-holomorphic polynomials for the case $s=-j$. 

 Returning to the zero-mode equation \eqref{zeroeq}, 
we first consider the case where only the top component of $\Psi$ is a non-zero function, which we assume to have the factorised form $R(\rho)\Sec(z_1,z_2)$. For this to be a zero-mode, the function $\Sec(z_1,z_2)$ 
 has to be annihilated by  $\lv_+$ and thus  holomorphic in $z_1,z_2$. It follows that $s=j$ in this case. Fixing $j$ and using   \eqref{holobasis},  we deduce the general form of the solution as 
\bee
\label{holosol}
\Psi(r,z_1,z_2) = \begin{pmatrix}
R_j(\rho)\sum_{m=-j}^ja_m z_1^{j-m}z_2^{j+m} \\
0\\
0 \\
0
\end{pmatrix}.
\eee
Inserting into \eqref{zeroeq} leads to the radial equation
\bee
\label{jq}
\left(\partial_\rho+\left(\frac 1 2 (p-1)-j\right)+ \left ( \frac 1 2 -j \right)\frac 1 \rho  -\frac {1}{2\rho(\rho+1)}\right)R_j(\rho)=0,
\eee
which has the general solution
\bee
\label{posq}
R_j(\rho)=c\frac{\rho^j}{\sqrt{\rho+1}}e^{\left(j-\frac{p-1}{2}\right)\rho},
\eee
for some constant $c\in\CC$.
This solution  is normalisable provided
\bee
j <\frac{p-1}{2} \Leftrightarrow 2j+1 < p, 
\eee
which can only happen if $p >1$.

To find solutions for the case $p<0$, we consider spinors $\Psi$ where only the second component is non-vanishing and of the form $\tilde{R}(\rho)\Sec(z_1,z_2)$. For this  to be a zero-mode, $F$  it has to be annihilated by $\lv_-$, so has to be anti-holomorphic. It follows that $s=-j$ in this case. Fixing $j$ and using \eqref{antiholobasis},  we deduce the general form of the solution as 
\bee
\label{antiholosol}
\Psi(r,z_1,z_2)= \begin{pmatrix}
0 \\
\tilde{R}_j(\rho)\sum_{m=-j}^j \tilde{a}_m\bar{z}_1^{j-m}\bar{z}_2^{j+m}\\
0 \\
0
\end{pmatrix}. 
\eee
Inserting into \eqref{zeroeq} 
leads to the radial equation
\bee
\left(\partial_\rho  -\left(\frac 1 2 (p+1)+j\right)   +\left(\frac 1 2-j\right)\frac 1 \rho  -\frac {1}{2\rho(\rho+1)}\right)\tilde{R}_j(\rho)=0.
\eee
This is the equation \eqref{jq} with $p$ replaced by $-p$. The general solution is therefore
\bee
\label{negq}
\tilde{R}_j(\rho)=\tilde{c}\frac{\rho^j}{\sqrt{\rho+1}}e^{\left(j+\frac{p+1}{2}\right)\rho},
\eee
for some $\tilde{c}\in \CC$. 
This solution  is normalisable provided
\bee
j < -\frac{p+1}{2} \Leftrightarrow 2j+1 <-p, 
\eee
which can only happen if $p < -1$.

 Concentrating on the case of $p>1$,  we count zero-modes by noting that the space of solutions for fixed $j$ has dimension $2j+1$. Again using our convention that $[p]$ is the largest integer {\em strictly} smaller than $p$ (so that [3]=2 etc), the  total  dimension of the space of zero modes is  
 \bee
 \text{dim ker}\Dslash_p=1 + 2 + \ldots [p]= \frac 1 2 [p]([p]+1), 
 \eee
 in agreement with Pope's formula \eqref{popedim}. We now  interpret this formula in terms of $SU(2)$ representations and Dirac monopoles.

 The action of $U\in SU(2)$ on the zero-modes is simply  via pull-back of the action of $U^{-1}$ on $z_1,z_2$. With the parametrisation of $U\in SU(2)$ in terms of complex numbers $a,b$ satisfying $|a|^2+|b|^2=1$ as in  \eqref{su2},  the action on \eqref{holosol} or \eqref{antiholosol} is 
\bee
U: \Psi(r,z_1,z_2)\mapsto \psi(r, \bar{b}z_1-\bar{a}z_2, az_1+bz_2). 
\eee
As reviewed in Appendix~\ref{harmonic}, the holomorphic (or antiholomorphic) homogeneous polynomials in $z_1,z_2$  of degree $2j$ form the $(2j+1)$-dimensional irreducible representation of $SU(2)$ under this action. This is  precisely   the action which we   encountered when studying the $SU(2)$ transformations of zero-modes of the twisted Dirac operator on the 2-sphere in \eqref{suupstairs}.  Thus we conclude that  the kernel of $\Dslash_p$ is  the sum of irreducible $SU(2)$ representation of dimension $\leq [p]$ or, equivalently,  the direct sum of the  kernels of the Dirac operators $\Dslash_{S^2,n}$ with $n=1,2,\ldots,[p]-1,[p]$.

To understand the latter interpretation better, recall that the Taub-NUT manifold may be thought of as a static Kaluza-Klein monopole of charge one \cite{Sorkin, GP}. In this geometrised description of the magnetic monopole, the $U(1)$ gauge symmetry is encoded in the $U(1)$-right action generated  by $\lv_3$. Functions, spinors or forms transforming non-trivially under this $U(1)$-action are electrically charged. For spinors, the operator
\bee
\hat N= 2i\hat \lv_3,
\eee
where $\hat \lv_3$ is defined in \eqref{hat3},
is the analogue of the `Chern-number operator' \eqref{ccoperator} introduced in the context of the twisted Dirac operator on the 2-sphere.  It 
 has integer eigenvalues $n$ which count the product of  the magnetic and electric charge. The eigenvalue is $n=2j+1  $ for the solution \eqref{holosol} in the case $p>1$ and is $n=-(2j + 1 ) $ for the solution \eqref{antiholosol} in the case $p<1$. As for the  Dirac operator $\Dslash_{S^2,n}$,  the absolute value of this  integer  gives the number of zero modes for a fixed $n$. Summing over all allowed values of $j$ (and hence $n$)   gives all zero modes. 

Reverting to the radial coordinate $r=\rho L$,  we observe that  the  radial function in \eqref{posq} and \eqref{negq} plays off exponential growth with coefficient $(2j+1)/(2L)$ against exponential decay with coefficient
$|p|/(2L)$. The exponential growth comes from the geometry of the Taub-NUT space while the decay comes entirely from the  auxiliary $\RR$-gauge field. The  effective length scale $2L/(|p|-2j-1)$ 
 plays a role analogous to that of $\Lambda$  in the solutions \eqref{massivesolutions} of the massive Dirac equation on $\RR^3$, but it  only has the correct sign if $|p| > 2j+1$. 

\begin{figure}[!ht]
\centering
\includegraphics[width=5truecm]{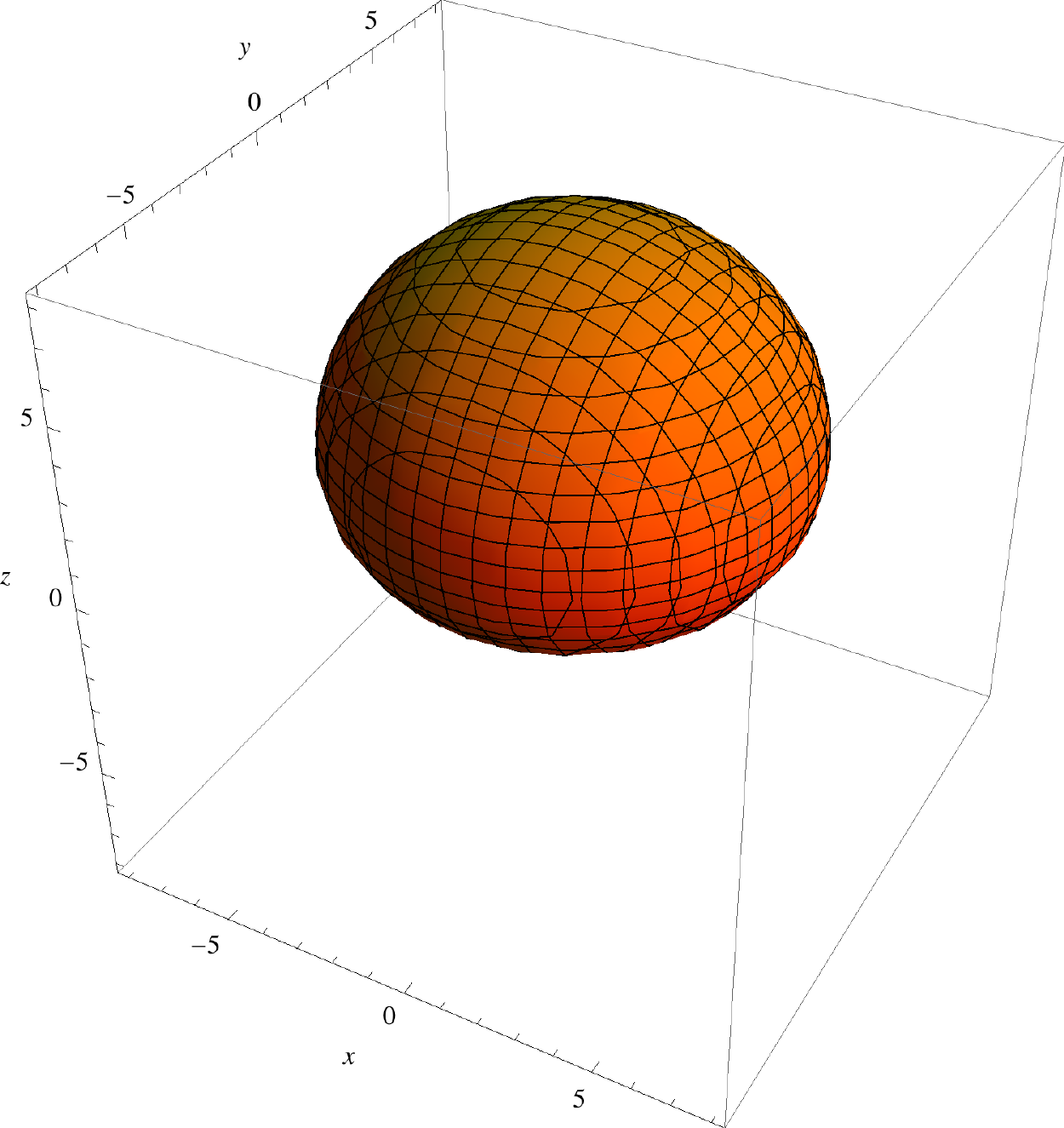}
\includegraphics[width=5truecm]{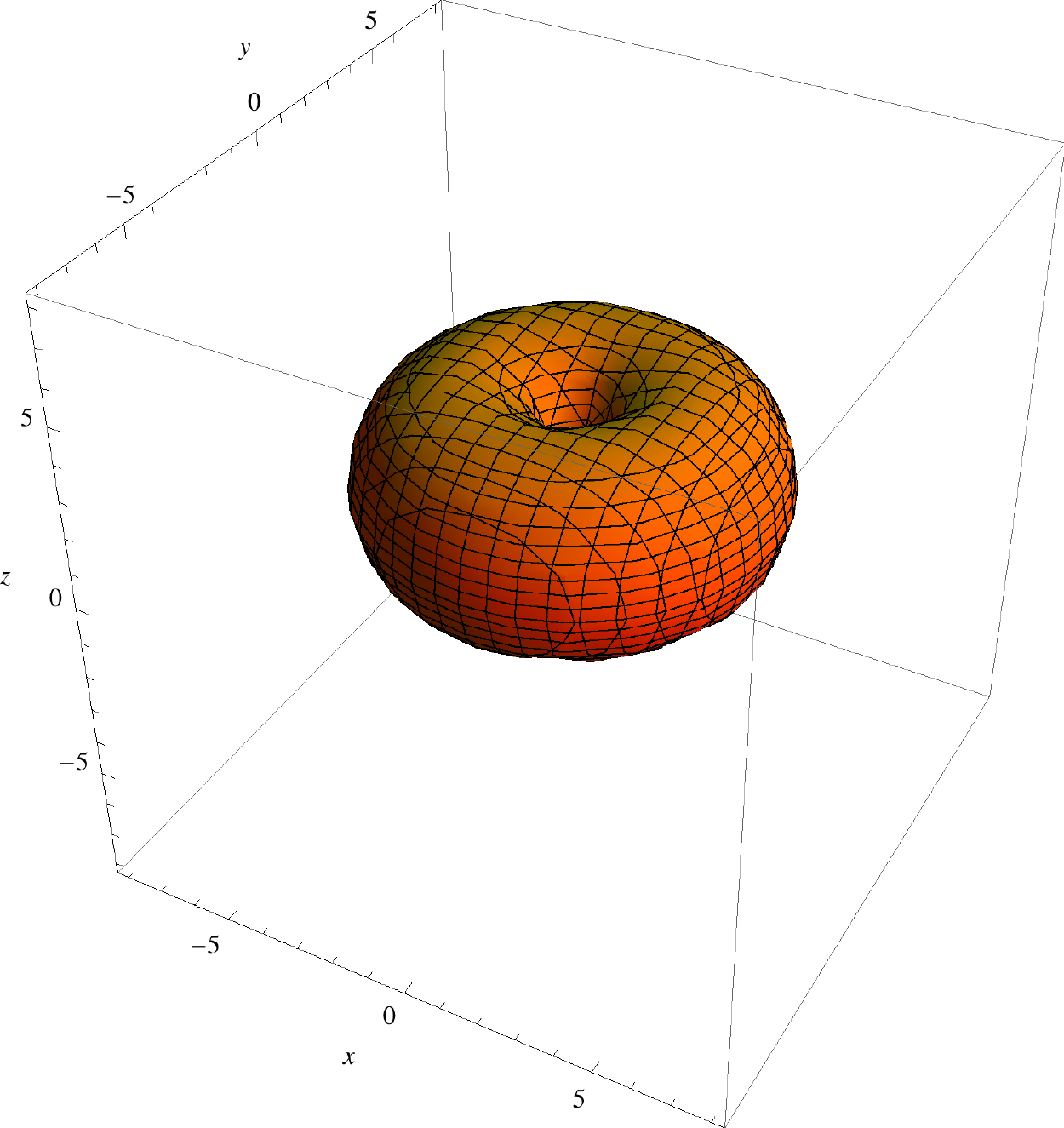}
\includegraphics[width=5truecm]{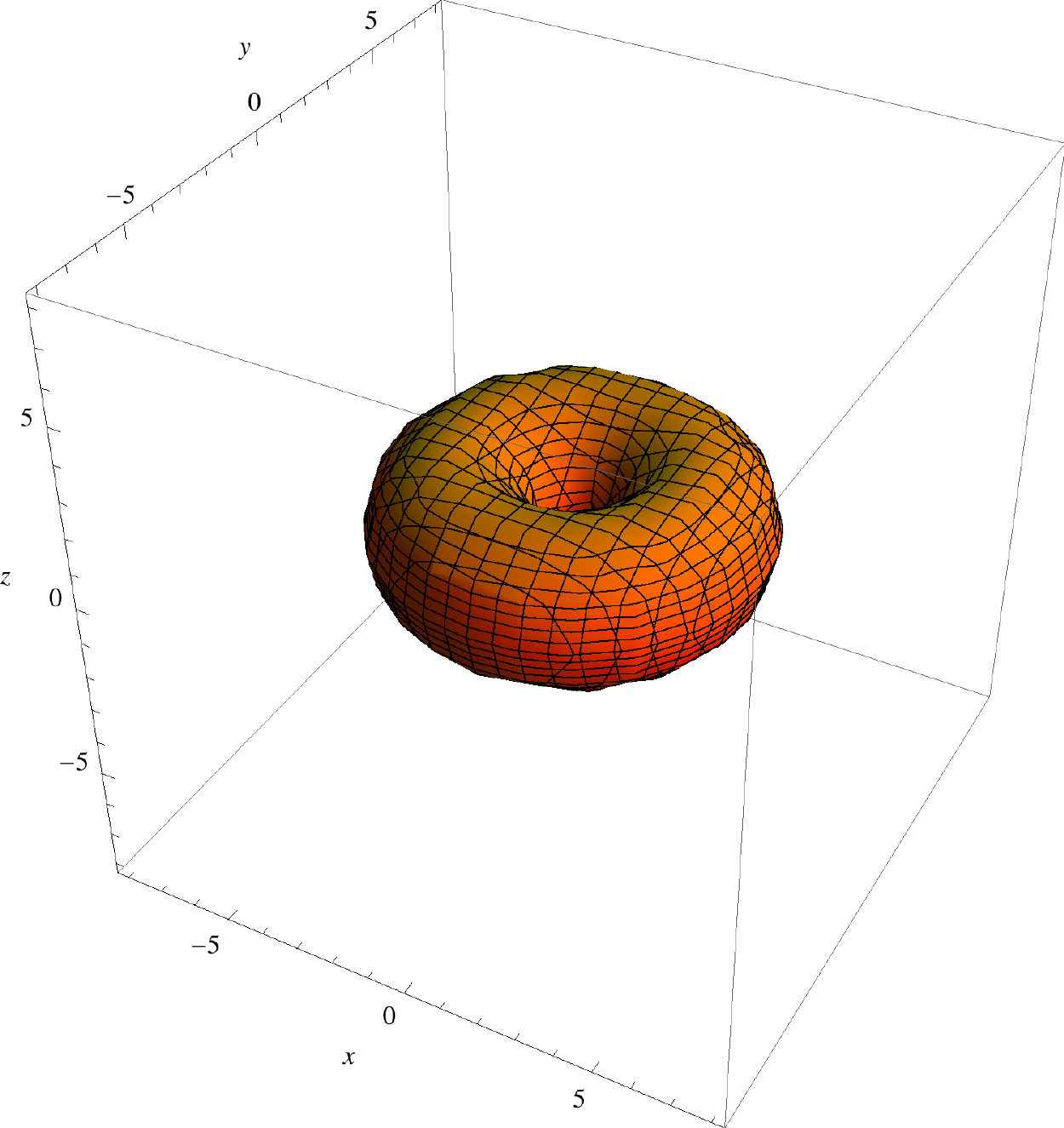}
\caption{Density contours of the squared zero-mode \eqref{probdens} for $j=4$  and $p=12$ and, from left to right, $m=-4,m=-2,m=0$ }
\label{genspinpic}
\end{figure}

To end our discussion of the zero-modes, we would like to point out that they define interesting geometrical shapes in 3-dimensional Euclidean space even though they are defined on the 4-dimensional Taub-NUT manifold. The reason is that their dependence on the $U(1)$ fibre of Taub-NUT (viewed as a circle-bundle over $\RR^3\setminus\{0\}$) is a pure phase. Thus, their square - which would 
give a probability distribution in a hypothetical quantum mechanical interpretation of   the  zero-modes  - only depends on 
the position in $\RR^3$, given by 
\bee
(x_1,x_2,x_3) = (r\sin\beta\cos\alpha, r\sin\beta\sin \alpha, r\cos\beta),
\eee
 see also our discussion of the Hopf fibration before \eqref{unitvector}. Focusing on $p>1$ and picking a term of fixed $m$ in the zero-mode \eqref{holosol}, we  obtain  the axially symmetric distribution
\bee
\label{probdens}
|\Psi|^2(x_1,x_2,x_3) \propto \frac{e^{(2j+1-p)\frac{r}{L} } } {r+L} (r-x_3)^{j+m}(r+x_3)^{j-m}.
\eee
For $-j <m <j$, it  vanishes along the entire $x_3$-axis. For $j=m$,  it is zero only for $x_3\geq 0$ while for $j=-m$ it vanishes for $x_3 \leq 0$. We show contour plots of typical zero-modes in Fig.~\ref{genspinpic}.

\section{Conclusion}
 
We end with  some  general observations and comments on our results.
Having understood the $SU(2)$ transformation properties of the zero-modes, it remains a puzzle why $SU(2)$ representations with a range of different spins are degenerate in the kernel of $\Dslash_p$. The degeneracy grows quadratically in the `quantum number' $[|p|]$ and is reminiscent of  generic energy eigenspaces for the Hamiltonian of the non-relativistic hydrogen atom and, closer to the current context, for  the Laplace  and the Dirac operator on the Taub-NUT space (not twisted by a connection).  In all cases,  the degeneracy can be understood in terms of an additional conserved vector operator - the quantum analogue of the Runge-Lenz vector \cite{Visinescu}. We have not investigated generalisations of this operator for the twisted Dirac operators studied here. In any case, an argument based on symmetry would not be entirely satisfactory since the index  of the operator is invariant under small changes of both the metric and the connection which would destroy any symmetry. For a topological degeneracy like the one studied here, one expects there to be a more robust reason. 

Our discussion could be extended and generalised to the multicentre Taub-NUT space, for which the dimension of the kernel of an appropriate Dirac operator was already given by Pope in \cite{Pope2} as the  dimension \eqref{popedim} times the number of centres. Other interesting four-manifolds with natural candidates for line bundles and connections are the Atiyah-Hitchin manifold, the complex projective plane with the  Fubini-Study metric as well  the Hitchin family of 4-manifolds which interpolates between them. All of these spaces are described in \cite{AMS}, where they are proposed as possible geometric models for elementary particles.

In the interpretation of the Taub-NUT manifold as a geometric model for the electron in \cite{AMS}, zero-modes of the Dirac operator were proposed as possible carriers of the spin 1/2 degrees of freedom of the electron. With the length scale  $L$ of the Taub-NUT manifold identified with the  classical electron radius as proposed in \cite{AMS},  
the  zero-modes  are  localised to the size of the classical electron radius. Focusing on positive $p$, our discussion also shows that the  kernel of  $\Dslash_p$ does indeed contain a normalisable doublet of spin 1/2 states, provided we  pick $p>2$. 
 To obtain spin {\em at most} 1/2, we need $p\leq 3$, but even with this choice we retain a spin 0 singlet as well. We have not been able to eliminate the spin 0 state by any natural condition. 
 
 However, we note that spin 1/2 states have one special property among all the zero-modes. By picking  $p=2$, the spin 1/2 doublet has the functional dependence
\bee
\label{spin12}
\sqrt{\frac{r}{r+L}} (a_{-1} z_1+ a_{1} z_2),
\eee 
which tends to  $SU(2)$ doublet states  in their standard form  $a_{-1} z_1+ a_{1} z_2$ as $r\rightarrow \infty$. Uniquely among the zero-modes, spin 1/2 states can be made to neither decay to zero nor blow up  at spatial infinity by a choice of $p$. With the same choice  $p=2$, the square  \eqref{probdens} of the spin 0 state is  exponentially localised at the origin, with characteristic size $L$. It is proportional to
\bee
\frac{ e^{-\frac  r L} } {r+L}. 
\eee
Borrowing supersymmetry jargon, the choice $p=2$ therefore gives a totally delocalised spin 1/2 `soul'  and an exponentially localised spin 0 `body'.

\noindent {\bf Acknowledgements} \; RJ thanks the School of Mathematical and  Computer Sciences at Heriot-Watt University for a Global Platform PhD Scholarship. BJS thanks Michael Singer for discussions about Dirac operators coupled to monopoles. 

\vspace{0.5cm}
\appendix

\section{Background and conventions}

\subsection{Parametrising $SU(2)$}
\label{parametrisations}
Our conventions and coordinates  in this paper are designed to be convenient for describing the Hopf map, harmonic analysis on $S^3$ and sections of powers of the hyperplane bundle over $S^2$. To achieve this,  we picked different  
 conventions from  those in  \cite{GM,MS,Schroers,AMS} which study closely related material. In particular, our $\mathfrak{su}(2)$ generators have the opposite sign of the ones used in those papers.  As a result, the left-invariant forms and vector fields change sign.  Our choice of Euler angles is also different. 

To parametrise   the group $SU(2)$,   we    use  the $\mathfrak{su}(2)$  generators 
\bee
\label{liegen}
t_j=-\frac{i}{2}\tau_j, \; j=1,2,3,
\eee
where $\tau_a$ are the Pauli matrices; the commutators are
$[t_i,t_j]=\epsilon_{ijk}t_k$.
We then  pararmetrise $h\in SU(2)$ in terms of Euler angles 
$\beta\in [0,\pi)$, $\alpha\in[0,2\pi)$ and $\gamma\in[0,4\pi)$
 as follows
\bee
\label{geuler}
h= e^{\alpha t_3}e^{\beta t_2 } e^{\gamma t_3}= \begin{pmatrix}
e^{-\frac i 2 (\gamma + \alpha)} \cos \frac{\beta}{2} & -e^{\frac i 2 (\gamma-\alpha)} 
\sin \frac{\beta}{2}\\
e^{\frac i 2 (\alpha-\gamma)} \sin \frac{\beta}{2} &
e^{\frac i 2 (\gamma + \alpha)} \cos \frac{\beta}{2} \end{pmatrix}.
\eee
We also use an alternative parametrisation  in terms of a complex unit vector  $(z_1,z_2)$  as 
\bee
\label{complexp}
h= \begin{pmatrix}
z_1  & -\bar{z}_2\\
z_2 &
\phantom{-} \bar{z}_1 \end{pmatrix},
\eee
with the constraint $ |z_1|^2 + |z_2|^2= 1$ understood. Comparing with \eqref{geuler}, we have
\bee
\label{zdeff}
z_1= e^{-\frac i 2 (\alpha+ \gamma)} 
\cos \frac{\beta}{2} , \quad 
z_2=
e^{\frac i 2 (\alpha - \gamma)} \sin \frac{\beta }{2}.
\eee

\subsection{Forms and  vector fields  on $SU(2)$}
\label{formsfields}

With $h\in SU(2)$ and the generators $t_j,\;  j=1,2,3$, defined in \eqref{liegen} we define the left-invariant 1-forms on $SU(2)$ via
\begin{equation}
\label{leftinvariant}
h^{-1}d h = \lf_1 t_1 + \lf_2 t_2 + \lf _3 t_3 \,.
\end{equation}
For the Euler angle parametrisation \eqref{geuler} we   compute to find
\begin{align}
 \lf_1 &=  \sin\gamma d\beta - \cos\gamma \sin\beta d\alpha,  \nonumber\\
 \lf_2 &=  \cos\gamma d\beta + \sin\gamma \sin\beta d\alpha, \nonumber\\
\lf_3 &=  d\gamma  + \cos\beta d\alpha.
\end{align}
These forms  satisfy $d\lf_i =- \frac{1} {2} \epsilon_{ijk}
 \lf_j \wedge \lf_k $.

 The dual vector fields $\lv_j$, $j=1,2,3$,  are left-invariant and generate the infinitesimal right-action
 \bee
\label{lvdef}
\lv_j: h\mapsto ht_j, \quad j=1,2,3.
\eee
Their commutators are 
\bee
[\lv_i,\lv_j]=\epsilon_{ijk} \lv_k.
\eee 
In the main text we  often use the combinations
\bee
\lv_+ = \lv_1+i\lv_2, \qquad \lv_-=\lv_1-i\lv_2,
\eee
which satisfy
\bee
[i\lv_3,\lv_\pm]=\pm \lv_\pm,
\eee
and therefore act as raising (+) and lowering (-) operators for $i\lv_3$.
In terms of Euler angles  we find 
\begin{align}
\label{lvectors}
 \lv_1 &= \phantom{-}\cot\beta \cos \gamma \partial_\gamma
           + \sin \gamma \partial_\beta
            -\frac {\cos \gamma }{ \sin\beta} \partial_\alpha,
\nonumber\\
\lv_2 &= -\cot\beta \sin \gamma \partial_\gamma
           + \cos \gamma \partial_\beta
            + \frac {\sin \gamma}{ \sin\beta} \partial_\alpha,
\nonumber \\
 \lv_3 &= \phantom{-}\partial_\gamma,
\end{align}
 so that 
 \begin{align}
\label{vfieldangles}
\lv_+= i e^{-i\gamma}\left(\partial_{\beta} + i \frac {1}{\sin \beta}
\partial_{\alpha}\ - i\frac{ \cos\beta}{\sin\beta} \partial_{\gamma}\right), \quad 
\lv_-=  -ie^{i\gamma}\left(\partial_{\beta} - i\frac{1}{\sin \beta} \partial_{\alpha}+ i \frac{\cos \beta}{\sin\beta }\partial_{\gamma}\right).
\end{align}
We also require   the left-invariant 1-forms and vector fields in complex notation. With \eqref{complexp}, we find
\bee
\lf_1+i\lf_2= 2i(z_1 dz_2 - z_2dz_1), \qquad  \lf_3= 2i(\bar{z}_1 dz_1 + \bar{z}_2dz_2).
\eee

To compute the dual vector fields in complex notation we use 
\bee
t_+=t_1+it_2= -i\begin{pmatrix}  0 &  1  \\ 0  & 0 \end{pmatrix}, \quad t_-=t_1-it_2= -i\begin{pmatrix}  0 &  0  \\ 1  & 0 \end{pmatrix}.
\eee
Then, from the rule \eqref{lvdef}
we have, for example.
\bee
\lv_+: \begin{pmatrix}
z_1  & -\bar{z}_2\\
z_2 &
\bar{z}_1 \end{pmatrix} \mapsto -i\begin{pmatrix}
z_1  & -\bar{z}_2\\
z_2 &
\bar{z}_1 \end{pmatrix}\begin{pmatrix}  0 &  1  \\ 0  & 0 \end{pmatrix}.
\eee
Evaluating, we find 
\begin{align}
\lv_+& =i (z_1\bar{\partial}_2 -z_2 \bar{\partial}_1), \nonumber  \\
\lv_-& =i (\bar{z}_2\partial_1 -\bar{z}_1 \partial_2), \nonumber \\
\lv_3& =\frac  i  2 ( \bar{z}_1\bar{\partial}_1 +\bar{z}_2 \bar{\partial}_2 - z_1 \partial_1 -z_2 \partial_2) .
\end{align}
One checks  that 
\bee
\lf_+(\lv_-)=\lf_-(\lv_+)=2, \quad \lf_3(\lv_3)=1,
\eee
with all other pairings vanishing.

Similarly, for left-generated  and right-invariant vector fields
\bee
\rv_i: h\mapsto -t_ih,
\eee
we define $\rv_\pm = \rv_1\pm i\rv_2$ and find
\begin{align}
\label{leftgenerators}
\rv_+& = i(z_2\partial_1 -\bar{z}_1\bar{\partial}_2), \nonumber \\
\rv_- & = i(z_1\partial_2 -\bar{z}_2\bar{\partial}_1), \nonumber \\
\rv_3& =\frac i 2  (z_1 \partial_1 - z_2 \partial_2 - \bar{z}_1\bar{\partial}_1 +\bar{z}_2 \bar{\partial}_2   ).
\end{align}
They satisfy $[\rv_i,\rv_j]=\epsilon_{ijk}\rv_k$  (and hence $[i\rv_3,\rv_\pm]=\pm \rv_\pm$)  and  commute with the right-generated vector fields $\lv_j$, $j=1,2,3$.   

\subsection{Harmonic analysis on $S^3$ in complex coordinates}
\label{harmonic}
The Laplace operator on $SU(2)$ acting on functions on $SU(2)$  can be written as 
\bee
\label{s3laplace}
\Delta_{S^3}= \lv_1^2 +\lv_2^2 + \lv_3^2 = \rv_1^2 +\rv_2^2 + \rv_3^2. 
\eee
It commutes with left- and right-generated vector fields, and its eigenspaces can therefore be decomposed into irreducible representations of $\mathfrak{su}(2)\oplus \mathfrak{su}(2)$, generated by $\lv_j$ and $\rv_j$, $j=1,2,3$. Here, we are only interested in the decomposition of  functions on $SU(2)$ into irreducible representations  under the $SU(2)$ left-action,  generated by $\rv_j$, $j=1,2,3$. Since these generators commute with $i\lv_3$ and $\Delta_{S^3}$, we can fix the eigenvalues of both $i\lv_3$ and $\Delta_{S^3}$.  We  now show how to obtain
the  irreducible representations under the $SU(2)$ actions in this way, using complex coordinates.

We use the trick of abandoning the constraint $|z_1|^2 + |z_2|^2$  and considering functions  defined on all of $\CC^2$, see \cite{Sternberg} for an analogous  treatment of the  Laplace operator on $S^2$. In order to obtain irreducible representations of $SU(2)$ we need to impose  the constraint that  the Laplace operator on $\CC^2\simeq \RR^4$
\bee
\label{4dlaplace}
\Box_4=  4 (\partial_1\bar{\partial}_1 +\partial_2\bar{\partial_2})
\eee
vanishes.

To see how and why this works,  we  define differential operators on  $\CC^2$
\bee
D=\frac 1 2 (z_1\partial_1 +z_2\partial_2), \qquad \bar{D}=\frac 1 2 (\bar{z}_1\bar{\partial}_1 +\bar{z}_2\bar{\partial}_2),
\eee
and observe   that both $D$ and $\bar{D}$ commute with $\rv_\pm, \rv_3$ and that 
\bee
i\lv_3 = D-\bar{D}.
\eee
We also find that 
\bee
\lv_+\lv_- = -4D\bar{D}-2 D + (|z_1|^2+|z_2|^2)(\partial_1\bar{\partial}_1 +\partial_2\bar{\partial_2}),
\eee
and therefore   have the identity 
\begin{align}
\Delta_{S^3}&  = X_+X_- +(D-\bar{D}) - (D-\bar{D})^2 \nonumber \\
& = - (D+\bar{D})^2 -(D+\bar{D}) +(|z_1|^2+|z_2|^2)(\partial_1\bar{\partial}_1 +\partial_2\bar{\partial_2}).
\end{align}
Defining
\bee
J=D+\bar{D},
\eee
we conclude  that
\bee
\Delta_{S^3}\Sec=-J(J+1)\Sec, \qquad \text{provided} \quad \Box_4\Sec =0.
\eee

Picking  half integers $N,\bar{N}\in \frac 1 2 \NN_0$ and   $ m,\bar{m}\in \frac 12 \ZZ$  in the range
\bee
m\in \{-N, -N+1,\ldots ,N-1, N\} ,\quad \bar{m} \in \{ -\bar{N}, \bar{N}+1,\ldots, \bar{N}-1,\bar{N}\},
\eee
and defining a monomial
\bee
F_{Nm\bar{N}\bar{m}}=z_1^{N-m}z_2^{N+m}\bar{z}_1^{\bar{N}+\bar{m}}\bar{z}_2^{\bar{N}-\bar{m}}, 
\eee
one checks that 
\bee
DF_{Nm\bar{N}\bar{m}}=NF_{Nm\bar{N}\bar{m}}, \quad \bar{D}F_{Nm\bar{N}\bar{m}}=\bar{N}F_{Nm\bar{N}\bar{m}}, \
\eee
and hence
\bee
JF_{Nm\bar{N}\bar{m}}= (N+\bar{N}) F_{Nm\bar{N}\bar{m}},  \qquad
 i\lv_3F_{Nm\bar{N}\bar{m}} =(N-\bar{N})F_{Nm\bar{N}\bar{m}}.
 \eee

We can now see that imposing  the annihilation by $\Box_4$   projects out an irreducible representation of $SU(2)$ as follows. 
We fix the eigenvalues $N$ and $\bar{N}$, and hence also $j:=N+\bar{N}$ and $s:=N-\bar{N}$. 
Then we write $P_{(N,\bar{N})}$ for the space of polynomials in $z_1,z_2,\bar{z}_1,\bar{z}_2$ with fixed values $N,\bar{N}$. Thus, $P_{(N,\bar{N})}$ has dimension
  $(2N+1)(2\bar{N}+1)$. It is easy to check that
\bee
\Box: P_{(N,\bar{N})} \rightarrow P_{(N-\frac 12 ,\bar{N}-\frac 1 2 )}
\eee
is surjective. As a result, the kernel  
has dimension 
\bee
d= (2N+1)(2\bar{N}+1)- 4 N\bar{N}=2(N+\bar{N})+1 =2j+1.
\eee
The monomial 
$F_{N N \bar{N} \bar{N}}$ is in this space , and is an eigenstate of $i\rv_3$: 
\bee
 i\rv_3 F_{NN\bar{N}\bar{N}} = (N +\bar{N})F_{NN\bar{N}\bar{N}} = jF_{NN\bar{N}\bar{N}} .
\eee
Acting with the lowering operator $\rv_-$ we generate the $(2j+1)$-dimensional irreducible representation of $SU(2)$, as claimed.

We are not going to give a basis for this space in the general case, but note two special cases which are used in the main text. When $s=j$, we have $\bar{N}=0, N=j$ and obtain the (non-normalised) holomorphic basis
\bee
\label{holobasis}
z_1^{j-m}z_2^{j+m}, \qquad m=-j,-j+1,\ldots,j-1,j,
\eee
with elements labelled by the eigenvalue $m$ of $i\rv_3$.
 When $s=-j$, we have $N=0,\bar{N}=j$ and obtain the (non-normalised) antiholomorphic basis
\bee
\label{antiholobasis}
\bar{z}_1^{j+m}\bar{z}_2^{j-m} , \qquad m=-j,-j+1,\ldots,j-1,j,
\eee
with elements again labelled by the eigenvalue  $m$ of $i\rv_3$.

\subsection{Lens spaces and the Hopf fibration}
\label{Hopfconventions}
Identifying $S^3$ with $SU(2)$, the Hopf map $S^3\rightarrow  S^2$ is  defined by taking the quotient of $SU(2)$ by a $U(1)$ right action. To make this concrete we  pick the torus generated by $t_3$  to define 
the right action   
\bee
\label{raction}
R(e^{i\delta}):h \mapsto h e^{\delta  t_3}, \quad \delta \in [0,4\pi).
\eee
 In terms of  Euler angles, this is simply the shift
$\gamma \mapsto \gamma + \delta$. In terms of the complex coordinates $(z_1,z_2)$, the map reads
\bee
\label{ractionz}
R(e^{i\delta}):(z_1,z_2)\mapsto (z_1e^{-i\frac \delta 2}, z_2e^{-i\frac \delta 2}).
\eee
The infinitesimal generator is the vector field  $\lv_3$ in \eqref{x3defined}.

We need to generalise our discussion to include  the Lens space $L(1,n)=S^3/\ZZ_n$, obtained from $S^3$ by the right action of  the cyclic group $\ZZ_n$, $n\neq 0$, whose generator acts via
\bee
\label{Lensid}
h\mapsto h e^{\frac{4\pi}{n}t_3}, \qquad (z_1,z_2)\mapsto (z_1e^{-i\frac {2\pi}{n}}, z_2e^{-i\frac{2\pi}  {n}}).
\eee
The $U(1)$ right-action is as in 
\eqref{raction} but with  $\delta\in [0,4\pi/n)$. As a result the associated   basis of the $U(1)$ Lie algebra is $n i/2$.
The vector field on $SU(2)$ generated by the $U(1)$ right-action is still $\lv_3$, but is now the push-forward
of  the $U(1)$ generator $n i/2$:
\bee
\label{lensinfini}
 R_*\left( n\frac{i}{2} \right)=\lv_3.
\eee

The Hopf map can be written concretely as a projection from $L(1,n)$ onto the unit  2-sphere inside the Lie algebra  $\mathfrak{su}(2)$. The following formula holds strictly only for $S^3$, but it makes sense for $L(1,n)$,   too,  since the image  is manifestly invariant under \eqref{Lensid}:
\bee
\pi: S^3 \rightarrow S^2\subset \mathfrak{su}(2), \quad h\mapsto ht_3h^{-1}.
\eee
In terms of the Euler angle parametrisation \eqref{geuler}, 
\bee 
\label{unitvector}
\pi(h) = (\sin\beta \cos \alpha) t_1 + (\sin\beta \sin \alpha)t_2  +  (\cos \beta )t_3, 
\eee
so that our choice of Euler angles induces  $(\beta,\alpha)$ as standard spherical polar coordinates on the 2-sphere.

We  introduce complex coordinates on $S^2$ by stereographic projection. 
Writing  $N$ for the `North Pole' $ (0,0,1)\in S^2$
and  $S$ for the `South Pole' $ (0,0,-1)\in S^2$, we define   
\bee
 U_N= S^2\setminus\{S\}, \qquad U_S = S^2\setminus\{N\}.
 \eee
Then, in terms the coordinates \eqref{unitvector},   stereographic projection  from the South Pole is 
\bee
\label{St1}
\text{St}: U_N \subset S^2 \rightarrow  \CC, \qquad (n_1,n_2,n_3)\mapsto z= \frac{n_1+in_2}{1+n_3},
\eee
and  stereographic projection  from the  North  Pole, followed by complex conjugation is
\bee
\label{St2}
\bar{\text{St}}: U_S \subset S^2 \rightarrow  \CC, \qquad (n_1,n_2,n_3)\mapsto \zeta= \frac{n_1-in_2}{1-n_3}.
\eee
Thus $\zeta = 1/z$ and  we observe that 
\bee 
\label{zdefinedd}
z=\frac{z_2}{z_1} = \tan\frac{\beta}{2} e^{i\alpha} , \qquad 
\zeta=\frac{z_1}{ z_2} =\cot\frac{\beta}{2} e^{-i\alpha}.
\eee
In other words, in complex coordinates, the Hopf map followed stereographic project from the South Pole is 
\bee
\text{St}\circ\pi: S^3 \rightarrow U_N, \quad (z_1,z_2)\mapsto z,
\eee
 while  the Hopf map followed by stereographic projection  from the  North  Pole and  complex conjugation is 
\bee
\bar{\text{St}}\circ \pi: S^3\rightarrow U_S, \quad   (z_1,z_2)\mapsto \zeta.
\eee
In our discussion we also require   local sections of the Hopf bundle in both complex coordinates and Euler angles. 
We use the same notation for both  and write, 
  on the northern patch, 
\begin{align}
\label{eNdefined}
s_N:  U_N\rightarrow S^3, \quad
 z \mapsto  \frac{1}{\sqrt{1+|z|^2}}(1,z), \qquad (\beta,\alpha) \mapsto e^{\alpha t_3} e^{\beta t_2} e^{-\alpha t_3}
\end{align}
and  on the southern patch
 \begin{align}
 \label{eSdefined}
s_S: U_S \rightarrow S^3, \quad
 \zeta\mapsto \frac{1}{\sqrt{1+|\zeta |^2}}(\zeta,1),\qquad
(\beta,\alpha) \mapsto e^{\alpha t_3} e^{\beta t_2} e^{\alpha t_3}.
\end{align}

\subsection{Associated line bundles and  their sections}
\label{bundles}
Our discussion in the main text frequently describes sections of line bundles associated to the Lens spaces in terms of equivariant functions
\bee
\label{equidef}
\Sec: L(1,n)\rightarrow \CC,
\eee
i.e., functions which satisfy
\bee
\label{equiv1}
\Sec(he^{\delta t_3})=e^{-i\frac{n}{2} \delta}\Sec(h), \quad \delta \in \left[0,\frac {4\pi}{n}\right],
\eee
or, in complex coordinates,
\bee
\label{equiv2}
\Sec( \lambda z_1, \lambda  z_2)= \lambda^n \Sec(z_1,z_2), 
\eee
where we wrote $\lambda= e^{-i  \delta/ 2 }$. 
In order to minimise notation, we   use  $h$  also for elements of  $L(1,n)$ here (rather than equivalence classes).  
Infinitesimally,  the equivariance condition  can be expressed as 
\bee
\label{sconstapp}
iX_3\Sec =\frac n 2 \Sec.
\eee

We can obtain local sections on the patches 
 $U_N$ and $U_S$   via pull-back with \eqref{eNdefined} and \eqref{eSdefined}:
\bee
\sec_N=s_N^*\Sec, \qquad  \sec_S=s_S^*\Sec.
\eee
Using \eqref{equiv2}
and 
\bee
\sec_N(z)= \Sec\left(\frac{1 }{\sqrt{q}}(1,z)\right),\quad 
\sec_S(z)=\Sec\left(\sqrt{ \frac{\bar{z}}z}\frac{1 }{\sqrt{q}}(1,z)\right),
\eee
 one deduces the patching condition
\begin{align}
\label{patchcond}
\sec_S=e^{-in\alpha}\sec_N =\left(\frac{\bar{z}} {z} \right)^{\frac n 2}\sec_N.
\end{align}

The line bundle associated to $L(1,n)$ is  often denoted as $H^n$,  the $n$th tensor power of  the hyperplane bundle $H$.  The latter is  the dual bundle of the tautological line bundle $L$ over 
$\CP^1$ whose   fibre over a point $\ell \in\CP^1$ is the line in $\CC^2$ defined by $\ell$:
\bee
L=\{(l,(w_1,w_2)\subset \CP_1\times \CC^2| (w_1,w_2)\in l\}.
\eee
 For the   hyperplane bundle $H$ over  $\CP^1$,  the fibre over a point  $\ell \in \CP^1$ is the dual space $\ell^*$. In the equivariant language \eqref{equiv2},  holomorphic sections of  $H^n$, $n\geq 0$,  can be written as homogeneous polynomials of degree $n$ in the variables $z_1,z_2$:
\bee
\Sec (z_1,z_2) = \sum_{k=0}^n a_k z_1^{n-k}z_2^k.
\eee
The space of all holomorphic sections can then be identified with the $(n+1)$-dimensional space of all such  polynomials.  As we shall check below, the Chern number of $H^n$ is $n$.

\subsection{Invariant connections and the Dirac monopole}
\label{mopoconventions}
 The magnetic monopole of charge $n\neq 0$ is the curvature of the rotationally invariant $U(1)$ connection on the Lens space $L(1,n)$. Using \eqref{lensinfini}, the requirement for a 1-form $\cA$ to be a connection 1-form on $L(1,n)$ is 
\bee
 \cA(X_3)= \frac{in}{2},
 \eee
 while `rotationally invariant' means invariant under the left-action of $SU(2)$ on $L(1,n)$. 
The  form 
\bee
\label{conn}
\gp=\frac{in}{2} \sigma_3=\frac{in}{2}\left(d \gamma+ \cos \beta \, d \alpha \right).
\eee
 satisfies  both these  requirements.  Its curvature  is 
\bee
F=d\gp = -\frac{in}{2} \sin\theta d\beta \wedge d \alpha,
\eee
which is the field of the Dirac magnetic monopole. 

We obtain the local gauge potentials  via pull-back with the local sections \eqref{eNdefined} and \eqref{eSdefined}:
\begin{align}
\label{Dm}
s^*_N\cA= A^n_N  = \frac{in}{2}(-1+\cos\beta) d\alpha,  
\quad s^*_S\cA= A^n_S = \frac{in}{2}(1+\cos\theta) d\alpha.
\end{align}
 The potentials are related by the $U(1)$ gauge transformation 
 \bee
 A_S^n = A_N^n + g_{SN} d g_{SN}^{-1}, \qquad   g_{SN}(\alpha) = e^{-in\alpha},
 \eee
 and satisfy  $F=dA_N^n=dA_S^n$.  The charge $n$ must be an integer by the Dirac quantisation condition and equals the Chern number of the bundle
 \bee
 \frac{i}{2\pi} \int_{S^2} F = n.
\eee 
 Since the potential $A^n_N$ is well defined on $U_N$ we rewrite it in terms of $z$ and $q$ as  
\bee
A_N^n =\frac {n}  {2q} (zd\bar{z} - \bar{z}dz ),
 \eee
Similarly, on $U_S$, we have 
\bee
A_S^n= 
\frac{n}{2} \frac{\zeta d\bar{\zeta}- \bar{\zeta} d\zeta } {1+|\zeta|^2}.
\eee
For the curvature we find
\bee
F=n(d z_1\wedge d\bar{z}_1 + d z_2\wedge  d\bar{z}_2) =  n  \frac{d z\wedge d\bar{z}  }   {(1+|z|^2)^2} = n\frac{d \zeta\wedge d\bar{\zeta}} {(1+|\zeta|^2)^2},
\eee
with the equalities holding wherever the  expressions are defined.

\subsection{Conventions related to the Dirac operator}
\label{Diracconventions}
We will use the following conventions when writing down the Dirac operator on a Riemannian manifold. Introducing and $n$-bein of 1-forms $e_1,\ldots,e_n$ so that the metric is 
\bee
ds^2= e_1^2 + \ldots + e_n^2,
\eee
we solve 
\bee
\label{spincon}
de_a + \omega_{a b } \wedge e_b=0,
\eee
for the spin connection 1-forms $\omega_{ab}= -\omega_{ba}$, $a,b  =1,\ldots, n$.
In terms of the   dual vector fields $E_a$ defined via
\bee
e_a(E_b) =\delta_{ab},
\eee
and $\gamma$-matrices satisfying 
\bee
\{\gamma_{a},\gamma_{b}\} = -2\delta_{ab}, 
\eee
the  spin connection is 
\bee
\label{Gammadef}
\Gamma = - \frac{1} {8} [\gamma_{a},\gamma_{b}] \omega^{ab}.
\eee
The  Dirac operator takes the form
\begin{align}
\label{metricdirac}
\Dslash  = \gamma^{c}\iota_{E_c}(d +\Gamma) =\gamma^{c}\left(E_{c} - \frac{1} {8} [\gamma_a,\gamma_{b}] \omega^{ab}_{c}\right),
\end{align}
where  $\omega^{ab}_c =
\omega^{ab}(E_{c})$, and indices are moved up or down for convenience.  
When we twist the bundle of spinors with an additional $U(1)$ bundle with  connection $A$, the twisted Dirac operator is
\begin{align}
\label{gendirac}
\Dslash_A = \gamma^{c}\iota_{E_c}(d + A + \Gamma) = \gamma^{c}\left(E_{c} +A_c- \frac{1} {8} [\gamma_{a},\gamma_{b}] \omega^{ab}_{c}\right).
\end{align}

\end{document}